\documentclass[reprint,prl,aps, floatfix]{revtex4-1}
\usepackage[utf8]{inputenc}
\usepackage[T1]{fontenc}
\usepackage{graphicx}
\usepackage{xcolor}
\usepackage{dcolumn}
\usepackage{bm}
\usepackage{amssymb}
\usepackage{braket}
\usepackage{color,amsmath}
\usepackage{comment}
\usepackage{gensymb}
\usepackage{soul,xcolor} 
\setstcolor{red} 
\usepackage{epstopdf}
\usepackage{hhline}
\usepackage{tabularx}
\usepackage{xspace}
\usepackage{layouts}
\usepackage{lineno}
\usepackage{xr} 
\usepackage{amsfonts}
\usepackage{bbold}
\usepackage{placeins}

\usepackage{booktabs}
\usepackage{colortbl}
\usepackage{float}

\newcolumntype{C}[1]{>{\centering\arraybackslash}m{#1}}
\newcolumntype{N}{@{}m{0pt}@{}}

\newcommand{\moire}{moir\'e\xspace}

\newcommand{\Tc}{T_c}
\newcommand{\Bpar}{B_\parallel}
\newcommand{\Bperp}{B_\perp}
\newcommand{\Idc}{I_{\mathrm{dc}}}

\begin{document}

\title{Incipient superconductivity and tunable Chern insulators in twisted bilayer-trilayer graphene}

\author{Derek Waleffe$^{1*}$} 
\author{Aryana Bhattacharyya$^{1*}$}
\author{Manish Kumar$^{1}$} 
\author{Eric Maginnis$^{2}$}
\author{Anna Okounkova$^{1}$}
\author{Tobias Faehndrich$^{5, 6}$}
\author{Kenji Watanabe$^{3}$}
\author{Takashi Taniguchi$^{4}$}
\author{Joshua Folk$^{5,6}$}
\author{Matthew Yankowitz$^{1,2\dagger}$}

\affiliation{$^{1}$Department of Physics, University of Washington, Seattle, Washington, 98195, USA}
\affiliation{$^{2}$Department of Materials Science and Engineering, University of Washington, Seattle, Washington, 98195, USA}
\affiliation{$^{3}$Research Center for Electronic and Optical Materials, National Institute for Materials Science, 1-1 Namiki, Tsukuba 305-0044, Japan}
\affiliation{$^{4}$Research Center for Materials Nanoarchitectonics, National Institute for Materials Science, 1-1 Namiki, Tsukuba 305-0044, Japan}
\affiliation{$^{5}$Department of Physics and Astronomy, University of British Columbia, Vancouver, British Columbia, V6T 1Z1, Canada}
\affiliation{$^{6}$Quantum Matter Institute, University of British Columbia, Vancouver, British Columbia, V6T 1Z1, Canada}

\affiliation{$^{*}$These authors contributed equally to this work.}
\affiliation{$^{\dagger}$myank@uw.edu (M.Y.)}

\begin{abstract}
Moir\'e superlattices assembled by twisting Bernal and rhombohedral multilayer graphene host a rich set of interaction-driven magnetic and topological states, yet superconductivity has not been observed in these systems except in proximity to a transition-metal dichalcogenide. Here we report incipient superconductivity and tunable Chern insulators in twisted bilayer-trilayer graphene encapsulated by hexagonal boron nitride. Across twist angles from $\theta=1.05^\circ$ to $1.50^\circ$, Chern insulators form at integer and fractional \moire fillings for electron doping, with Chern numbers up to $|C|=3$ set by twist angle and tuned by doping. At $\theta=1.18^\circ$, a symmetry-broken metallic region forms for hole doping and hosts a trivial insulator at band filling $\nu=-2$. Displacement field alone drives this insulator into a pocket of incipient superconductivity with a sharp transition, a well-defined critical current, and a critical temperature that peaks near the insulating boundary, although the resistance does not fall to zero. In-plane magnetic field expands the pocket, which persists to more than four times the weak-coupling Pauli limit, and stabilizes a second pocket in which Fraunhofer-like modulation of the critical current signals phase-coherent pairing. Twisted bilayer-trilayer graphene thus offers a single gate-tunable system in which pairing can be interfaced with Chern insulators whose topology is itself an adjustable parameter.
\end{abstract}

\maketitle

Flat bands in van der Waals (vdW) materials provide a setting in which superconductivity, magnetism, and nontrivial topology can emerge within a single gate-tunable platform~\cite{Andrei2020_GrapheneBilayersTwist,Mak2022_SemiconductorMoireMaterials,Nuckolls2024_MoireMaterials,Cao2026_FQAH}. Across many vdW systems built from graphene and transition metal dichalcogenides, superconductivity often emerges adjacent to a symmetry-broken phase, raising questions about the association between magnetic ordering and pairing. The quantum geometry of flat bands can also contribute to the superfluid stiffness~\cite{Peotta2015_TopologicalFlatBands,Tian2023_DiracFlatBandSC}, while their topology supports integer- and fractionally quantized anomalous Hall effects~\cite{Sharpe2019_TBGFerromagnetism,Serlin2020_QAH,Cai2023_FQAH,Park2023_FQAH,Lu2024}. Systems where superconductivity and zero-field Chern insulators appear within the same phase diagram allow studies of the interplay among strong correlations, pairing, and topology, and open possibilities for engineering topological superconductivity at interfaces between these states. However, only a small subset of known vdW systems currently meets these conditions~\cite{Choi2025,Kumar2025_dual,Stepanov2021_ChernInsulatorsTBG,He2025_HofstadterTBG,Xu2025_UnconventionalSCFQAH}.

\begin{figure*}[t]
\includegraphics[width=\textwidth]{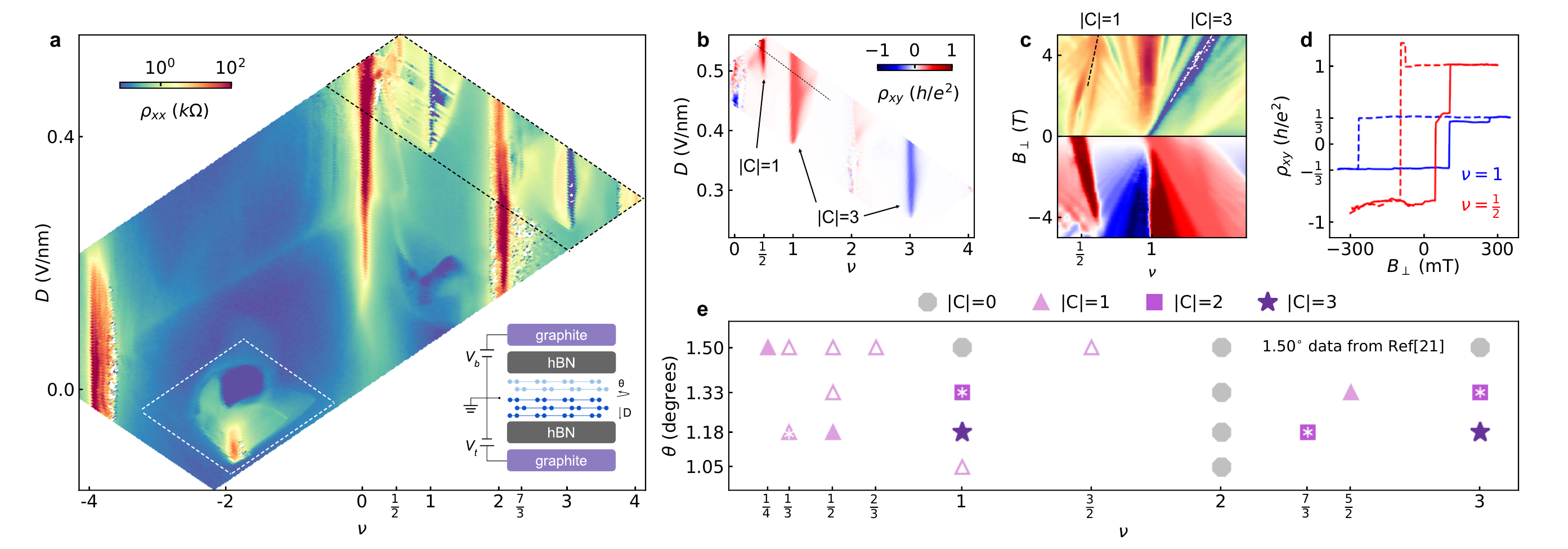} 
\caption{\textbf{Integer quantum anomalous Hall states in a device with $\mathbf{\boldsymbol{\theta}=1.18^\circ}$.} 
\textbf{a}, Map of the longitudinal resistivity $\rho_{xx}$ as a function of $\nu$ and $D$ taken at base temperature and zero magnetic field. (inset) Schematic of the dual-gated twisted bilayer-trilayer graphene device.
\textbf{b}, Map of the antisymmetrized Hall resistance $\rho_{xy}$ at $B= \pm100$~mT in the region denoted by the black dashed box in \textbf{a}. \textbf{c}, Landau fan diagram of symmetrized $\rho_{xx}$ (top) and antisymmetrized $\rho_{xy}$ (bottom) taken at $V_{t}=9.05$~V sweeping $V_b$ along the black dashed line in \textbf{b}. The black and white dashed lines indicate St\v{r}eda slopes corresponding to $C=1$ at $\nu=1/2$ and $C=3$ at $\nu=1$. Color scales are the same as in \textbf{a} and \textbf{b}.
\textbf{d}, Traces of $\rho_{xy}$ taken at $\nu=1$ and $1/2$ as $B$ is swept back and forth.
\textbf{e}, Chern numbers of correlated insulating states at different $\nu$ for devices of varying twist angle. Zero-field states denoted with a solid-filled shape and finite-field states denoted with an unfilled outline. The asterisks indicate states where the Chern number is ambiguous between two integer values, or between an integer and fractional value.
}
\label{fig:1}
\end{figure*}

Twisting graphene to form a moir\'e superlattice has provided a means for engineering topological flat bands. Systems with Bernal or rhombohedral multilayer graphene as the constituent components are particularly promising, as they exhibit a rich array of quantum anomalous Hall states with tunable Chern numbers~\cite{Polshyn2020,Chen2021_TMBG,He2021_TMBG,Waters2024_tMN,Su2025,Liu2025_HighChernTRG,Dong2025_HighChernFCI,Li2025_HighChernInsulators,Wang2026_QAHTwistedFlatbands,Chen2026_HighChernTRG,Wang2026_HighChernTRG,Dong2026_tRTBGPhaseTransitions}. However, these systems have not been reported to host superconductivity, except in the case of twisted double-bilayer graphene on a WSe$_2$ substrate~\cite{Su2023_TDBGWSe2}. This absence is especially curious since the Bernal bilayer or rhombohedral multilayer components themselves exhibit superconductivity~\cite{Zhou2021_RTGSC,Zhou2022_BBG}. Whether pairing is intrinsically disfavored in twisted multilayers, or has simply not been found under the right conditions, remains an open question.

Here we show that twisted Bernal bilayer-trilayer graphene hosts incipient superconductivity within the same gate-tunable phase diagram as its Chern insulators. We find that correlated insulators are common at integer and fractional \moire fillings across six devices spanning twist angles from $0.72^\circ$ to $1.89^\circ$. We focus on a device with $\theta=1.18^\circ$, in which the states at $\nu=\frac{1}{2}$, $1$, and $3$ are quantized at zero magnetic field. The same device hosts a correlated insulator at $\nu=-2$ within a symmetry-broken metallic region. Raising the displacement field at fixed filling carries the system out of this insulator and into a pocket of incipient superconductivity, with a maximum $\Tc\approx160$~mK neighboring the insulating boundary. Such a transition driven by band tuning has not been seen in other graphene systems, but is reminiscent of the electrically driven superconductor--insulator transition observed in twisted WSe$_2$~\cite{Xia2025_TwistedBilayerWSe2}. 

\medskip\noindent\textbf{Chern insulators at integer and fractional fillings}

Twisted Bernal bilayer-trilayer graphene is assembled from a single graphene flake containing adjacent bilayer and trilayer regions, which are separated and restacked with a controlled twist angle $\theta$ (see Supplementary Materials). The heterostructure is encapsulated in hexagonal boron nitride between top and bottom graphite gates, so that carrier density $n$ and displacement field $D$ can be set independently (inset of Fig.~\ref{fig:1}a). We express charge carrier density $n$ in terms of the \moire filling factor $\nu=4n/n_s$, where $n_s$ is the doping required to fill the four-fold degenerate \moire bands (see Supplementary Materials). Figure~\ref{fig:1}a shows the longitudinal resistivity $\rho_{xx}$ of device D1 ($\theta=1.18^\circ$) over a wide gate range at $B=0$ and at the nominal base temperature of our dilution refrigerators (see Supplementary Materials). Band insulators appear at charge neutrality and at $\nu=\pm4$. On the electron-doped side at large $D$, symmetry-broken halos surround $\nu=1$ and $\nu=3$ and a correlated insulator forms at $\nu=2$, following the pattern established in other twisted Bernal multilayers~\cite{Waters2024_tMN}. At low $D$, an independent resistive halo instead surrounds $\nu=-2$ for hole doping. We use ``halo'' throughout to denote the region of elevated resistance associated with an isospin-polarized metal.

\begin{figure*}[t]
\includegraphics[width=\textwidth]{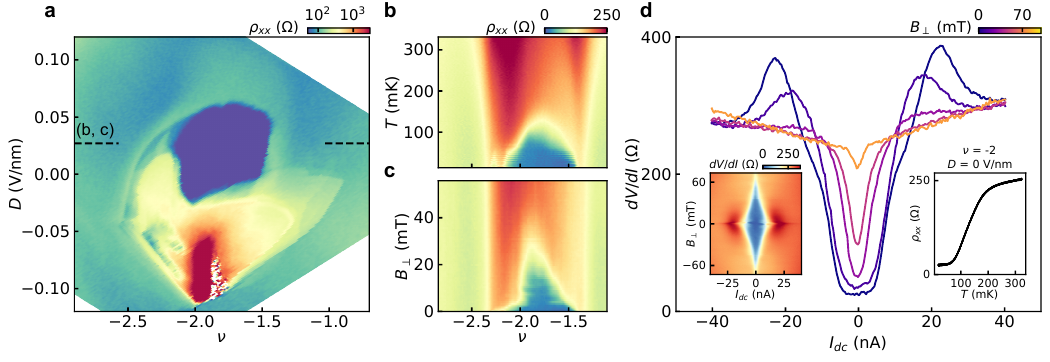} 
\caption{\textbf{Correlated insulator and incipient superconductivity at $\boldsymbol{\nu=-2}$.} \textbf{a}, Zoomed in map of $\rho_{xx}$ from the region of the white dashed box in Fig.~\ref{fig:1}a. 
\textbf{b}, Measurement of $\rho_{xx}$ versus $\nu$ and $T$ with fixed $D=0.027$~V/nm (corresponding to the black dashed line in \textbf{a}).
\textbf{c}, The same measurement versus $\Bperp$.
\textbf{d}, Measurement of d$V$/d$I$ versus $I_{dc}$ at selected values of $\Bperp$ taken at $\nu=-2$ and $D=0$. (left inset) Map of d$V$/d$I$ versus $I_{dc}$ and $\Bperp$. (right inset) Measurement of $\rho_{xx}$ versus $T$.
}
\label{fig:2}
\end{figure*}

The topological states arise in the high-$D$ electron-doped region outlined in Fig.~\ref{fig:1}a. A map of the Hall resistivity $\rho_{xy}$ in this region (Fig.~\ref{fig:1}b), antisymmetrized at $B=\pm100$~mT, shows enhancements at $\nu=1$ and $\nu=3$ that coincide with suppressions of $\rho_{xx}$, together with a sharp feature at $\nu=1/2$ where higher-resolution maps also reveal a corresponding resistance minimum (Fig.~\ref{fig:S_fractional}). Line traces taken at fixed $V_t$ and $B=0$ (Fig.~\ref{fig:S_addfans}) show $\rho_{xy}$ reaching approximately $h/e^2$ at $\nu=1/2$ and $h/3e^2$ at $\nu=1$ and $\nu=3$, with $\rho_{xx}$ suppressed to near zero at each filling, corresponding to Chern numbers $|C|=1$ and $|C|=3$. A weaker dip in $\rho_{xx}$ and peak in $\rho_{xy}$ also appear at $\nu=7/3$ (Figs.~\ref{fig:S_fractional}~\&~\ref{fig:S_addfans}).

Landau fans confirm the topological nature of the states. At $\nu=1/2$ and $\nu=1$ the Chern insulators disperse with out-of-plane magnetic field, following the appropriate St\v{r}eda trajectories (Fig.~\ref{fig:1}c). At $\nu=3$ a $C=+3$ state is additionally stabilized by magnetic field alongside the $C=-3$ state already present at $B=0$ (Fig.~\ref{fig:S_addfans}). Both $\nu=1$ and $\nu=1/2$ trace closed hysteresis loops with $\rho_{xy}$ quantized at $B=0$ (Fig.~\ref{fig:1}d). The loop at $\nu=1/2$ is well quantized when trained at $B>0$ but not at $B<0$; the origin of this training asymmetry is unresolved, but may result from twist angle inhomogeneity.

The Chern insulators at odd integer fillings have previously been attributed to symmetry-broken states in which both spin and valley are polarized~\cite{Polshyn2020,Chen2021_TMBG,He2021_TMBG,Waters2024_tMN}. In device D1, $\nu=1$ and $\nu=3$ carry the same $|C|=3$, consistent with each corresponding to one filled or empty flavor of a $C=3$ \moire band. The integer-quantized states at fractional filling most likely form due to a spontaneously enlarged unit cell commensurate with the \moire lattice. The resulting state is, in effect, a \moire-driven topological electronic crystal, with the Chern number set by the details of the folded band~\cite{Su2025}. A prior report on a $\theta=1.50^{\circ}$ device~\cite{Su2025} identified such states at $\nu=1/4$ at zero field and at $\nu=1/3$, $1/2$, $2/3$, and $3/2$ in modest field. The $\nu=1/2$ state in device D1 is a zero-field member of the same family, consistent with a doubled unit cell. We also observe a weak feature corresponding to a state at $\nu=1/3$ (Fig.~\ref{fig:S_fractional}). The feature at $\nu=\frac{7}{3}$ is more ambiguous. A Hall conductance of $\frac{7}{3}\,e^2/h$ has recently been reported in twisted Bernal bilayer-rhombohedral tetralayer graphene and attributed to a fractional Chern insulator~\cite{Li2025_HighChernInsulators}, but the corresponding Hall resistivity of $3h/7e^2$ differs from the $h/2e^2$ of an integer $C=2$ state by only 14\%, and the feature in our device is not developed well enough to distinguish the two.

Figure~\ref{fig:1}e collects every Chern insulator we identify across multiple devices spanning a range of twist angles, together with those reported at $\theta\approx1.5^\circ$ in Ref.~\cite{Su2025}. States which are not fully developed are identified by their Středa slopes in Landau fan diagrams (see Table \ref{tab:chern_numbers}). The Chern number ranges from 1 to 3 across integer and fractional fillings, and at a given filling it can shift with twist angle: $\nu=1$ carries $|C|=3$ at $1.18^\circ$ (device D1) and $C=0$ at $1.50^\circ$~\cite{Su2025}. Landau fan diagrams indicate a state with $|C|=2$ at $1.33^\circ$ (device D2, Figs.~\ref{fig:S_D2fans} and~\ref{fig:S_D2VTI}), while at $1.05^\circ$ we identify a state consistent with $|C|=1$ (device D3, Fig.~\ref{fig:S_1degree}). Topologically trivial correlated insulators, appearing most prominently at $\nu=2$, arise in five out of the six devices measured here and persist over a wider range of angles than the Chern states (Fig.~\ref{fig:S_gatemaps}).

\begin{figure*}[t]
\includegraphics[width=0.9\textwidth]{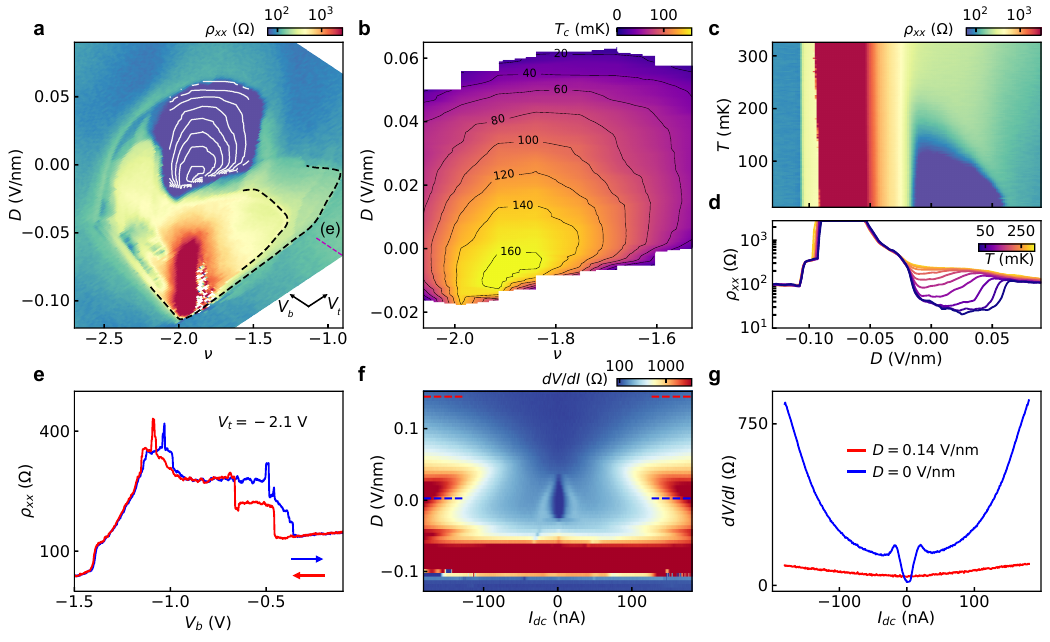} 
\caption{\textbf{Transition between a correlated insulator and a low-resistance pocket driven by $\boldsymbol{D}$.} 
\textbf{a}, The same map as in Fig.~\ref{fig:2}a with black dashed curves marking hysteretic features and white curves showing contour plots of approximately constant $T_c$.
\textbf{b}, Color heat map showing an interpolated map of $T_c$ versus $\nu$ and $D$. 
\textbf{c}, Measurement of $\rho_{xx}$ versus $D$ and $T$ with fixed $\nu=-1.91$.
\textbf{d}, Selected line traces from \textbf{c}.
\textbf{e}, Measurement of $\rho_{xx}$ as $V_b$ is swept back and forth. 
\textbf{f}, Map of d$V$/d$I$ versus $I_{dc}$ and $D$ at fixed $\nu=-2$.
\textbf{g}, Line traces of d$V$/d$I$ versus $I_{dc}$ from \textbf{f} taken at $D=0$ and $D=0.14$~V/nm.
}
\label{fig:3}
\end{figure*}

\medskip\noindent\textbf{Incipient superconductivity from a correlated insulator}

We now turn to the hole-doped side of the phase diagram and the resistive halo surrounding $\nu=-2$, outlined by the white dashed box in Fig.~\ref{fig:1}a and mapped at higher resolution in Fig.~\ref{fig:2}a. At negative $D$ the halo surrounds a correlated insulator at $\nu=-2$, flanked on both sides by metallic regions with elevated resistance compared to the surrounding portions of the map. These boundaries track lines of nearly constant gate voltage, indicating layer-screening effects~\cite{Kolar2025}. Quantum oscillations on either side of the insulator show a reduced degeneracy consistent with partial isospin polarization, and Landau fans identify the insulator as being topologically trivial ($C=0$; Fig.~\ref{fig:S_HaloLandauFans}).

Increasing $D$ at fixed $\nu=-2$ drives the system out of the correlated insulator and into a region where $\rho_{xx}$ is instead strongly suppressed. This low-resistance pocket shows features characteristic of a gate-tunable superconducting transition. Along a line of fixed $D=27$~mV/nm cutting through this region, $\rho_{xx}$ is sharply suppressed below a critical temperature, forming a characteristic dome in the $\nu$--$T$ plane (Fig.~\ref{fig:2}b). It shows similar behavior below a perpendicular critical field, forming a matching dome in the $\nu$--$\Bperp$ plane (Fig.~\ref{fig:2}c), with a critical field of several tens of millitesla. The critical temperature, taken as the point where $\rho_{xx}$ falls to half its normal-state value (see Supplementary Materials), reaches a maximal value of $\Tc\approx160$~mK. The differential resistance ($dV/dI$) is flat at small dc current ($I_{dc}$) and rises sharply at a critical current $I_c$ of order tens of nanoamperes. The critical current shrinks with $\Bperp$ and closes at $B_{c\perp}$, forming a characteristic diamond structure (Fig.~\ref{fig:2}d and its left inset; see also Fig.~\ref{fig:S_SCdVdI}).

\begin{figure*}[t]
\includegraphics[width=\textwidth]{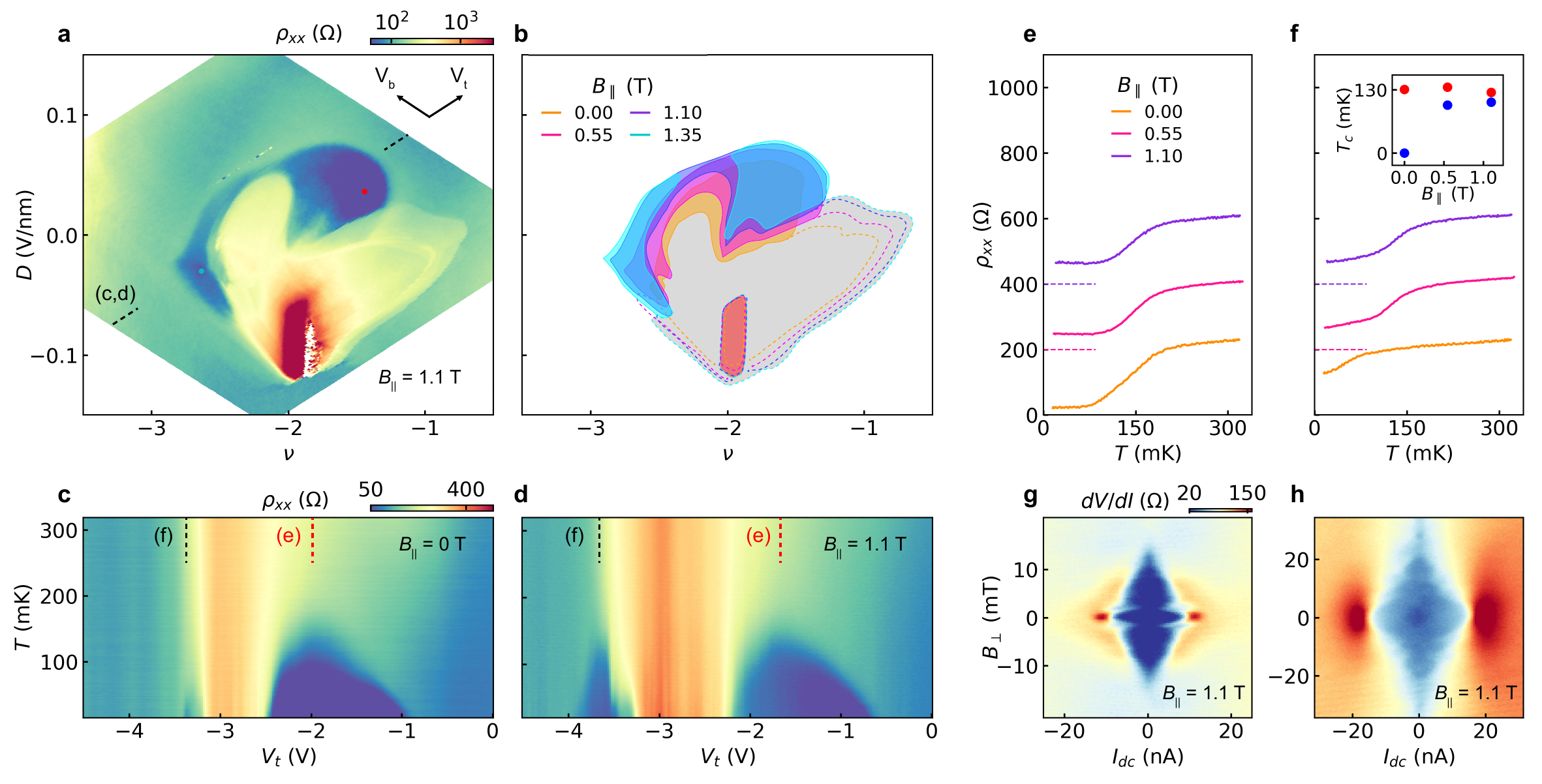} 
\caption{\textbf{Evolution with in-plane magnetic field.} 
\textbf{a}, Map of $\rho_{xx}$ versus $\nu$ and $D$ taken at $\Bpar=1.1$~T.
\textbf{b}, Schematic of the evolution of the halo region for selected values of $\Bpar$. The gray region denotes metallic states, with the boundary of the halo indicated by the dashed curves. The red region denotes the correlated insulator. The solid colored regions denote low-resistance pockets.
\textbf{c}, Measurement of $\rho_{xx}$ versus $V_t$ and $T$ with fixed $V_b=-1.85$~V at $\Bpar=0$ (along the direction indicated by the black dashed line in \textbf{a}).
\textbf{d}, The same map taken with $\Bpar=1.1$~T.
\textbf{e}, Measurements of $\rho_{xx}$ versus $T$ for selected values of $\Bpar$, taken at the position indicated by the corresponding dashed line in \textbf{c-d}. Curves are offset vertically for clarity, with zero denoted by the dashed lines.
\textbf{f}, Corresponding measurements taken at the dashed line labeled \textbf{f}. (inset) $\Tc$ versus $\Bpar$ from the measurements in \textbf{e-f}.
\textbf{g}, Map of d$V$/d$I$ versus $I_{dc}$ and $\Bperp$ taken at $\Bpar=1.1$~T at the position of the red dot in \textbf{a}.
\textbf{h}, The same map taken at the cyan dot in \textbf{a}.
}
\label{fig:4}
\end{figure*}

Despite otherwise exhibiting the typical hallmarks of superconductivity, the resistance in this pocket does not fall to zero. The right inset of Fig.~\ref{fig:2}d shows a representative trace taken at $\nu=-2$ and $D=0$, in which $\rho_{xx}$ drops sharply below $\approx160$~mK but eventually levels off at a finite value. The low-temperature behavior is not uniform across the pocket, but in no case does it reach zero. At some locations $\rho_{xx}$ plateaus over an extended range of temperature down to base, while at others it is still decreasing down to the lowest measured temperature (Fig.~\ref{fig:S_SC1DRT}). The same low-resistance pocket appears in measurements from a second pair of voltage probes elsewhere in the Hall bar (Fig.~\ref{fig:S_D1contactpair2}). We also see similar behavior over four thermal cycles spanning three different dilution refrigerators, with comparable boundaries in gate space and residual resistance (Fig.~\ref{fig:S_TandBdepHalo}). The finite resistance could potentially arise due to mesoscopic disorder, for instance from twist-angle inhomogeneity that forms regions of suppressed pairing between the voltage probes. Another possibility is that the state is an ``anomalous metal'' of the kind recently identified in rhombohedral graphene, where finite low-temperature resistance appears alongside zero-resistance superconductivity in a \moire-less system~\cite{Okounkova2026_AM, KKS2019}.

Taking the largest measured $B_{c\perp}\approx80$~mT gives an orbital Ginzburg--Landau coherence length $\xi_{\mathrm{GL}}=\sqrt{\Phi_0/2\pi B_{c\perp}}\approx64$~nm, where $\Phi_0=h/2e$, several times the 12~nm \moire wavelength at $\theta=1.18^\circ$. We refer to the state as an incipient superconductor since it shows the characteristic critical temperature, critical field, and critical current of a superconducting transition, but its resistance does not vanish. Hints of Fraunhofer-type interference oscillations are weakly resolved in this pocket at $\Bpar=0$, but appear clearly in the field-stabilized pocket discussed below (Fig.~\ref{fig:4}h). These oscillations provide strong evidence of phase coherence associated with a paired condensate.

\medskip\noindent\textbf{Maximal $\Tc$ near the insulating boundary}

The low-resistance pocket lies near both the correlated insulator at low $D$ and the polarization boundary of the halo, and the variation of $\Tc$ across gate space distinguishes which of the two is associated with the most robust pairing. We measure $\rho_{xx}$ versus $T$ and $D$ along lines of fixed $\nu$ spanning the pocket (Fig.~\ref{fig:3}c; see Supplementary Materials and Fig.~\ref{fig:S_TcFits}~\&~\ref{fig:S_TvsDJun21}), extract $\Tc$ at each point, and interpolate to produce the map in Fig.~\ref{fig:3}b. At every $\nu$, $\Tc$ is largest very near the low-$D$ edge of the pocket, and along that edge the maximum is concentrated in a narrow range of density around $\nu=-2$. Contour lines show $\Tc$ falling in every direction away from this region. Overlaid on the resistance map of the halo (Fig.~\ref{fig:3}a), the low-$D$ edge where $\Tc$ peaks faces the correlated insulator across a narrow metallic strip (Figs.~\ref{fig:3}c,d), and the range of $\nu$ over which $\Tc$ is maximal coincides with the range over which the insulator forms.

Portions of the halo boundary also exhibit hysteresis upon sweeping $n$ or $D$, suggestive of a first-order phase transition (Fig.~\ref{fig:3}e and Fig.~\ref{fig:S_Hysteresis}). In some regions we see consecutive hysteretic transitions, although we cannot tell whether these represent separate symmetry-broken phases or result from spatial inhomogeneity. The black dashed lines in Fig.~\ref{fig:3}a indicate these hysteretic boundaries, which lie away from the region of maximal $\Tc$.

Transport within the metallic portion of the halo is also nonlinear. Figure~\ref{fig:3}f maps $dV/dI$ versus $\Idc$ and $D$ at fixed $\nu=-2$, cutting through both the insulator and the low-resistance pocket. Within the pocket, d$V$/d$I$ does not level off once $\Idc$ exceeds $I_c$ as is typical, but instead continues to rise up to the largest current applied (Fig.~\ref{fig:3}g). This behavior spans the range of $D$ occupied by the pocket and disappears outside the halo, where transport is nearly linear. The same behavior appears along lines of fixed $D$ (Fig.~\ref{fig:S_dVdIconstD}). Further work is needed to assess the nature of the halo region, including the possibility of additional broken symmetries (e.g., translational or rotational) beyond the apparent partial isospin polarization.

\medskip\noindent\textbf{Field-stabilized pairing and Pauli-limit violation}

Figure~\ref{fig:4}a shows $\rho_{xx}$ around $\nu=-2$ at an in-plane magnetic field of $\Bpar=1.1$~T. We find that $\Bpar$ enlarges the halo without destroying the low-resistance pocket. Notably, a second region of sharply suppressed $\rho_{xx}$ forms along the leftmost edge of the halo, growing out of a tail of slightly reduced resistance that, at $B=0$, is already connected to the zero-field pocket (Fig.~\ref{fig:2}a). Figure~\ref{fig:4}b traces this evolution schematically (see Fig. \ref{fig:S_BparHaloEvolution}). The zero-field pocket expands slightly and shifts toward higher $\nu$ and $D$ up to $\Bpar\approx1.35$~T. The low-resistance tail grows rapidly, and eventually the region connecting them is pinched off by an expansion of the metallic region inside the halo, leaving two isolated pockets. The insulator and the halo as a whole also expand into the surrounding unpolarized metal as $\Bpar$ increases.

Temperature-dependent cuts provide further insights into this behavior. Figures~\ref{fig:4}c,d show $\rho_{xx}$ maps versus $T$ and $V_t$ at fixed $V_b=-1.85$~V, taken with $\Bpar=0$ and $1.1$~T, respectively. The dome corresponding to the zero-field pocket expands slightly in $V_t$ with increasing $\Bpar$, while a second dome at more negative $V_t$ appears and grows rapidly. Figures~\ref{fig:4}e,f show $\rho_{xx}$ versus $T$ at optimal doping in each pocket at $\Bpar=0$, $0.55$, and $1.1$~T, and the inset of Fig.~\ref{fig:4}f plots the extracted $\Tc$. In the zero-field pocket $\Tc$ is roughly unchanged across this range. In the field-stabilized pocket, $\Tc$ rises steeply between $0$ and $0.5$~T and then plateaus, converging on the same value as the zero-field pocket by $1.1$~T.

The zero-field pocket survives far beyond the Pauli limit. For a spin-singlet BCS superconductor, the paramagnetic Pauli limit is $B_P \approx 1.25\,k_B T_c/\mu_B$, where $k_B$ is the Boltzmann constant, $\mu_B$ is the Bohr magneton, and assuming a $g$-factor of 2. This corresponds to $B_P\approx0.30$~T for $\Tc\approx160$~mK. The pocket persists to at least $\Bpar=1.35$~T (Fig.~\ref{fig:S_BparHaloEvolution}d), more than four times $B_P$, and since $\Tc$ has not begun to fall at the highest field where it was measured, the true in-plane critical field lies beyond our range. For the field-stabilized pocket the Pauli limit is undefined at $\Bpar=0$, but a conservative bound follows from the largest $\Tc$ observed at any $\Bpar$, and on that basis the violation is also at least fourfold. Both pockets sit in a partially isospin-polarized metal, with the halo and pockets expanding into the surrounding metal as $\Bpar$ is raised. Together with the absence of the anomalous Hall effect, this behavior is consistent with a (partially) spin-polarized phase gaining Zeeman energy over an unpolarized phase, and supports a spin-triplet interpretation for pairing.

Figures~\ref{fig:4}g,h map $dV/dI$ versus $\Bperp$ and $\Idc$ at $\Bpar=1.1$~T in the zero-field and field-stabilized pockets. Both show a critical-current diamond characteristic of a superconducting state. The field-stabilized pocket additionally shows a Fraunhofer-like pattern of side lobes in $I_c(\Bperp)$ (Fig.~\ref{fig:4}h). Taking the lobe spacing $\Delta \Bperp\approx10$~mT gives an effective flux area $A=\Phi_0/\Delta \Bperp\approx0.2~\mu\mathrm{m}^2$. This is far smaller than the area enclosed between the voltage probes, so the modulation reflects a localized weak link within the current path rather than coherence across the full measured region. Its presence nonetheless provides strong evidence for phase-coherent pairing.

\medskip\noindent\textbf{Discussion}

The displacement-field-driven evolution from the $\nu=-2$ insulator into the paired state is reminiscent of recent observations in twisted WSe$_2$, where an electric field drives a superconductor--insulator transition at fixed commensurate filling~\cite{Xia2025_TwistedBilayerWSe2}. Subsequent studies on twisted WSe$_2$ away from commensurate filling support antiferromagnetic correlations in the neighboring phase and find the strongest superconductivity near its boundary~\cite{Xia2026_MottTransitionWSe2,Guo2026_tWSe2AngleEvolution}. In twisted Bernal bilayer-trilayer graphene, the surrounding phase is instead a symmetry-broken metal that we infer to be partially spin-polarized. The highest $\Tc$ occurs near the commensurate insulator within this region, concentrated over the same range of $\nu$ as the insulator and decreasing away from the insulating boundary, whereas the observed hysteretic boundaries of the halo lie elsewhere in gate space. Pairing in the zero-field pocket therefore tracks the commensurate insulator rather than the polarization transition at the edge of the halo. The order responsible for the insulator, and the origin of the enhanced $\Tc$ near its boundary, remain to be determined.

Incipient superconductivity in twisted Bernal bilayer-trilayer graphene has so far appeared only near $1.2^\circ$. A device at $\theta\approx1.05^\circ$ shows a similar correlated insulator and halo at $\nu=-2$ without a pocket (Fig.~\ref{fig:S_1degree}), but was only measured down to 1.5~K, well above the critical temperature in device D1. Twisted monolayer-bilayer graphene shows closely similar features at low twist angle, likewise without superconductivity so far~\cite{Chen2021_TMBG}. The halo is therefore stable over a range of angles and across the twisted Bernal family, while the conditions under which it also supports pairing remain to be identified. Further investigation in highly homogeneous twisted bilayer-trilayer devices as a function of angle will provide key insights into the relationship between the correlated insulator at $\nu=-2$ and superconductivity.

In summary, twisted Bernal bilayer-trilayer graphene hosts Cooper pairing and Chern insulators in the same electrically tunable phase diagram. Although both states also appear in crystalline rhombohedral graphene~\cite{Choi2025} and magic-angle twisted bilayer graphene aligned with boron nitride~\cite{Stepanov2021_ChernInsulatorsTBG,He2025_HofstadterTBG,Serlin2020_QAH}, twisted Bernal multilayers are distinguished from these other systems by the breadth of the topology already available in them, with $|C|$ up to 3 at zero field and a Chern number that depends on twist angle and gating. The Chern number therefore becomes a parameter that can be varied within a single device, opening a route to proximity experiments in which the topology on one side of a junction is varied in situ.

\medskip\noindent\textit{Note added.} During the preparation of this manuscript we became aware of a related result reporting superconductivity and Chern insulating states in twisted rhombohedral graphene multilayers~\cite{Huo2026_SCandHCItwR}.

\section*{Acknowledgments}
This work was primarily supported by University of Washington Molecular Engineering Materials Center, a U.S. National Science Foundation Materials Research Science and Engineering Center (DMR-2308979). Device fabrication was supported by the National Science Foundation (NSF) CAREER award no. DMR-2041972. A.B. acknowledges support from the University of Washington Mary Gates Endowment for Students and the Smith Endowment as well as the Washington Space Grant Consortium (NASA-80NSSC20M0104). Experiments at the University of British Columbia were undertaken with support from the Natural Sciences and Engineering Research Council of Canada; the Canada Foundation for Innovation; the Canadian Institute for Advanced Research; the Max Planck-UBC-UTokyo Centre for Quantum Materials and the Canada First Research Excellence Fund, Quantum Materials and Future Technologies Program; and the European Research Council (ERC) under the European Union's Horizon 2020 research and innovation programme, Grant Agreement No. 951541. K.W. and T.T. acknowledge support from the JSPS KAKENHI (Grant Numbers 21H05233 and 23H02052) and World Premier International Research Center Initiative (WPI), MEXT, Japan. This work made use of shared fabrication facilities at UW provided by NSF MRSEC 2308979.

\section{Author Contributions} 
D.W., A.B., M.K., and E.M. fabricated the samples; D.W. and A.B. led the measurements at UW with assistance from M.K. and E.M.; measurements at UBC in the lab of J.F. were performed by D.W. and A.B. with assistance from T.F. and A.O.; K.W. and T.T. provided the hBN crystals; D.W., A.B. and M.Y. wrote the manuscript with input from all authors; M.Y. supervised the project. 

\section*{Competing interests}
The authors declare no competing interests.

\section*{Additional Information}
Correspondence and requests for materials should be addressed to Matthew Yankowitz.

\section*{Data Availability}
Source data are available for this paper. All other data that support the findings of this study are available from the corresponding author upon request.


\newpage

\clearpage
\onecolumngrid
\setcounter{figure}{0}
\setcounter{equation}{0}
\renewcommand{\thefigure}{S\arabic{figure}}
\renewcommand{\thetable}{S\arabic{table}}

\begin{center}
\textbf{\large Supplementary Materials for:}\\[0.3em]
\textbf{\large Incipient superconductivity and tunable Chern insulators in twisted bilayer-trilayer graphene}
\end{center}

\section*{Methods}

\textbf{Device fabrication.} Graphite and hexagonal boron nitride (hBN) crystals were mechanically exfoliated and searched using an optical microscope. Graphene flakes with both bilayer and trilayer regions were identified by carefully counting monolayer and bilayer steps in the flake, indicated by changes in optical contrast. The desired domains were then isolated using a resist-free local anodic oxidation nanolithography process. All van der Waals heterostructures were assembled using standard dry-transfer techniques with a polycarbonate (PC) film on a polydimethylsiloxane (PDMS) stamp. Devices were assembled top down, as shown in the inset of Fig.~\ref{fig:1}a, and deposited onto an SiO$_2$ substrate. Standard reactive ion etching (CHF$_3$/O$_2$) and e-beam lithography fabrication processes were then used to define and contact the dual-gated Hall bar device. All lithography patterns were written on a poly(methyl methacrylate) (PMMA) resist layer and all devices were electrically connected using 7/70~nm of Cr/Au.

\begin{figure*}[h]
\includegraphics[width=0.3\textwidth]{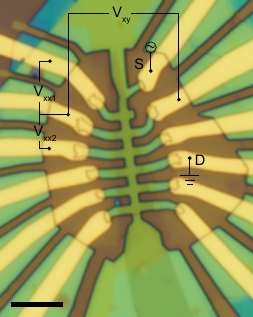} 
\caption{\textbf{Primary contact configuration for Device D1 ($\mathbf{\boldsymbol{\theta}=1.18^\circ}$)}. Scale bar is 5 $\mu$m.
}
\label{fig:S_Micrograph}
\end{figure*}

\textbf{Transport measurements.} Transport measurements on device D1 ($\theta\approx1.18^\circ$) were carried out in four thermal cycles: first in a Bluefors LD dilution refrigerator with a one-axis superconducting magnet, then in a Bluefors LD dilution refrigerator equipped with a three-axis superconducting vector magnet, again in the single-axis system, and finally in a second Bluefors LD dilution refrigerator equipped with a three-axis superconducting vector magnet and a fast sample exchange (FSE) system. The nominal base mixing chamber temperatures of the three systems are $T=10$~mK, $25$~mK, and $8$~mK. In the first thermal cycle the mixing chamber sensor read $T=50$~mK at base. We report the temperatures recorded by the factory-supplied sensor in each cycle without further correction, and note that data from the first and third cycles, taken in the same system with the same measurement configuration, are very similar. Four-terminal lock-in measurements were performed using a source current $I\leq10$~nA and a frequency less than 50~Hz, chosen to capture sensitive transport features while minimizing electronic noise. All $\rho_{xx}$ and $dV/dI$ measurements are presented after scaling the raw resistance measurements by the ratio of the Hall bar width to the spacing between voltage probes. We also applied a global back gate voltage between $15$~V and $-20$~V to the Si substrate to minimize contact issues. Figure~\ref{fig:S_Micrograph} shows an optical micrograph of device D1 and outlines the primary contact configuration used for the transport measurements. Transport measurements on both $\theta\approx1.33^\circ$ devices (D2 and D4) were also carried out in the Bluefors LD dilution refrigerator with a one-axis superconducting magnet. The mixing chamber temperature was constant around $T=10$~mK and the measurement setup was the same as described above. Additional measurements on D3 and all other samples were performed at or above 1.5~K in a Cryomagnetics variable temperature insert or a modified Janis G-M cryocooler.

Top and bottom graphite gates were used to independently tune the charge carrier density $n$ and displacement field $D$ in each sample. Conversions from gate voltages to $n$ and $D$ are given by $n=(C_{b}V_{b}+C_{t}V_{t})/e$ and $D=(C_{t}V_{t}-C_{b}V_{b})/2\epsilon_0$, where $C_{t}$ and $C_{b}$ are the top and bottom gate capacitance per unit area, $e$ is the elementary charge, and $\epsilon_0$ is the vacuum permittivity. The gate capacitances were estimated from fitting the slopes of quantum Hall states as they evolved with $\Bperp$.

Several gate maps and Landau fan diagrams were symmetrized or antisymmetrized to minimize mixing between the $\rho_{xx}$ and $\rho_{xy}$ signals. Symmetrized $\rho_{xx}$ data is given by $\rho_{xx}=(\rho_{xx}(B>0)+\rho_{xx}(B<0))/2$ and antisymmetrized $\rho_{xy}$ is given by $\rho_{xy}=(\rho_{xy}(B>0)-\rho_{xy}(B<0))/2$. In all measurements of the superconducting states versus $\Bperp$, the value of $\Bperp$ is adjusted such that $\rho_{xx}$ measurements are symmetric about $B=0$~T. This adjustment corrects against any trapped flux in the superconducting magnet coil. 

\textbf{Determination of twist angle.} The \moire band filling factor was defined by $\nu=n/(n_s/4)$, where $n_s$ is the carrier density required to fully fill a four-fold degenerate \moire band. The twist angle $\theta$ is related to $n_s$ by $n_s=8\theta^2/\sqrt{3}a^2$, where $a=0.246$~nm is the lattice constant of graphene. For each device, we estimate a twist angle based on the experimentally determined carrier density required to fully fill all four \moire bands. This can be done after fitting for the hBN capacitances using the known St\v{r}eda slopes of quantum Hall states in Landau fans of voltage versus out-of-plane magnetic field. In several cases, the twist angle was further verified by fitting the sequence of Brown-Zak magnetoconductance oscillations extracted from a Landau fan, serving as an independent check.

\begin{figure*}[h]
\includegraphics[width=0.95\textwidth]{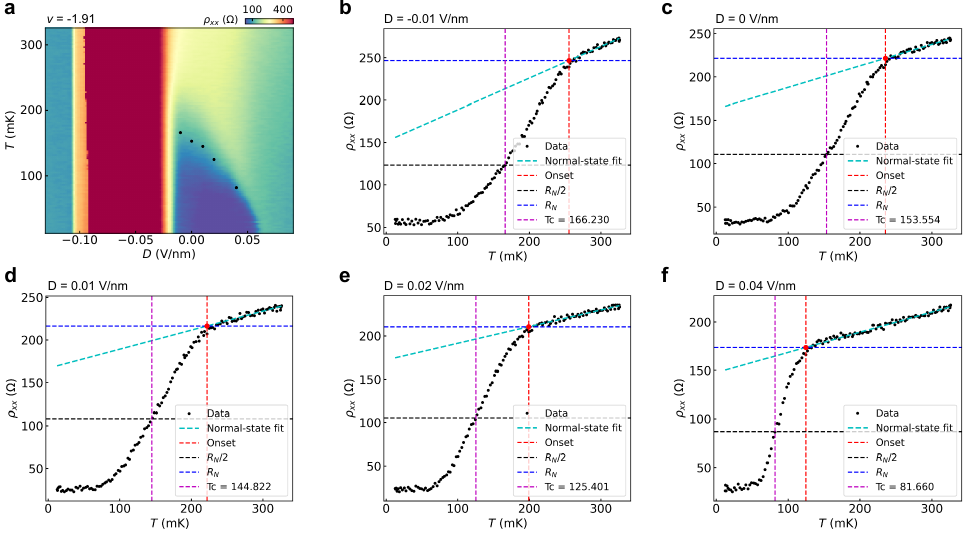} 
\caption{\textbf{Example $\boldsymbol{T_c}$ analysis}. \textbf{a}, $\rho_{xx}$ versus $T$ and $D$ taken at constant $\nu = -1.91$ across the low resistance pocket of Device D1. Black dots correspond to the $T_c$ values determined in $\textbf{b-f}$. \textbf{b}, $T_c$ analysis at $D = -0.01$~V/nm. \textbf{c}, $D = 0$~V/nm. \textbf{d}, $D = 0.01$~V/nm. \textbf{e}, $D = 0.02$~V/nm. \textbf{f}, $D = 0.04$~V/nm.}
\label{fig:S_TcFits}
\end{figure*}

\begin{figure*}[h]
\includegraphics[width=0.95\textwidth]{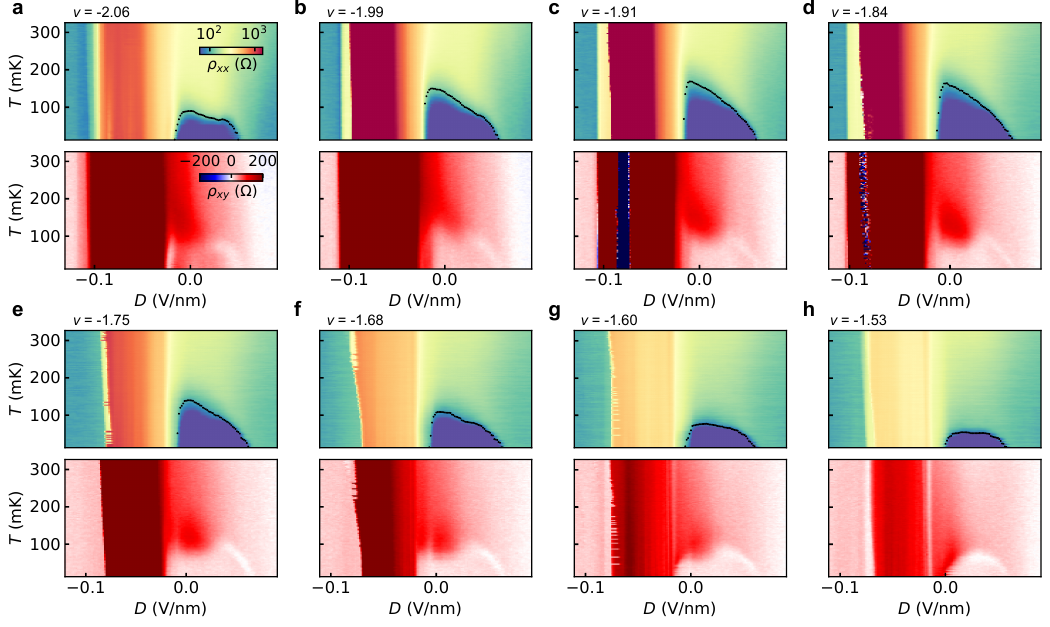} 
\caption{\textbf{$\boldsymbol{\rho_{xx}}$ versus $\boldsymbol{T}$ and $\boldsymbol{D}$ in the low resistance pocket}. \textbf{a}, (top) $\rho_{xx}$ versus $T$ and $D$ taken at constant $\nu = -2.06$ in the low resistance pocket of device D1. Black dots show $T_{c50\%}$, defined as the temperature at which $\rho_{xx}$ reaches $50\%$ of the normal state resistance. (bottom) Corresponding $\rho_{xy}$. \textbf{b}, Same for $\nu = -1.99$. \textbf{c}, $\nu = -1.91$. \textbf{d}, $\nu = -1.84$. \textbf{e}, $\nu = -1.75$. \textbf{f}, $\nu = -1.68$. \textbf{g}, $\nu = -1.60$. \textbf{h}, $\nu = -1.53$.
}
\label{fig:S_TvsDJun21}
\end{figure*}

\textbf{Critical temperature analysis.} In Figs.~\ref{fig:2},~\ref{fig:3},~\ref{fig:4} and Fig.~\ref{fig:S_TvsDJun21} we report on the critical temperature of the superconducting pockets. We define $\Tc$ as the temperature at which the resistance falls to $50\%$ of the normal state resistance $R_n$. $R_n$ is determined by fitting a line to the resistance at temperatures above $\Tc$ and locating the resistance where the data departs from the linear fit. Departure is defined as when the measured resistance deviates from the fit by more than 3 times the standard deviation of the residual of the fit. $R_n$ is assigned as the resistance given by the fit at this temperature. Example fits are shown in Fig.~\ref{fig:S_TcFits}. The qualitative features of the $\Tc$ map in Fig.~\ref{fig:3}b are not affected by choosing $30\%$, $50\%$, or $90\%$ of the normal state resistance as the definition of $\Tc$.

\begin{figure*}[h]
\includegraphics[width=0.95\textwidth]{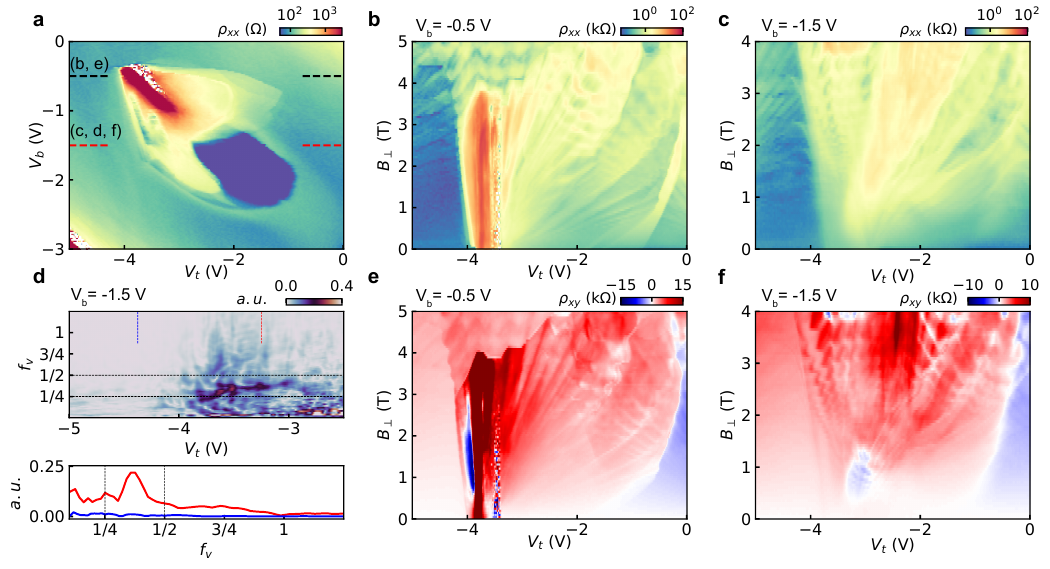} 
\caption{\textbf{Landau fans through the correlated halo}. \textbf{a}, Map of $\rho_{xx}$ versus $V_b$ and $V_t$ in the region surrounding the correlated halo at $\nu = -2$ (Device D1). Dashed lines indicate where the measurements in \textbf{b-f} are taken. \textbf{b}, Landau fan of $\rho_{xx}$ versus $B$ and $V_t$ taken at $V_b=-0.5$~V, along the black dashed line in \textbf{a}. \textbf{c}, Same as \textbf{b}, taken at $V_b=-1.5$~V along the red dashed line in \textbf{a}. \textbf{d}, (top) Fast Fourier transform (FFT) of $\rho_{xx}$($\frac{1}{B}$), taken on the data in \textbf{c}. (bottom) Linecuts taken along the blue ($V_t=-4.38$~V) and red ($V_t=-3.25$~V) dashed lines from (top). \textbf{e}, $\rho_{xy}$ corresponding to \textbf{b}. \textbf{f}, $\rho_{xy}$ corresponding to \textbf{c}.
}
\label{fig:S_HaloLandauFans}
\end{figure*}

\textbf{Fermiology analysis.} In Fig.~\ref{fig:S_HaloLandauFans}d, we take a fast Fourier transform (FFT) of $\rho_{xx}$($\frac{1}{B}$) in order to get information about the Fermi surface. As a part of this procedure we first fit and subtract a fifth order polynomial from the raw $\rho_{xx}$$(B)$ data. The subtracted data is then interpolated onto a regular grid in order to perform the Fourier transform. We additionally normalize the raw frequencies by the Luttinger volume, $n(\frac{h}{e})$. The normalized frequency $f_\nu$ corresponds to the fraction of the Fermi surface enclosed by a cyclotron orbit in momentum space, and therefore allows the Fermi surface degeneracy to be inferred. For the analysis in Fig.~\ref{fig:S_HaloLandauFans}d we use the magnetic field range of $B_\perp=1$~T to $B_\perp=4$~T, taken from the Landau fan in Fig.~\ref{fig:S_HaloLandauFans}c. 

\begin{figure*}[h]
\includegraphics[width=0.95\textwidth]{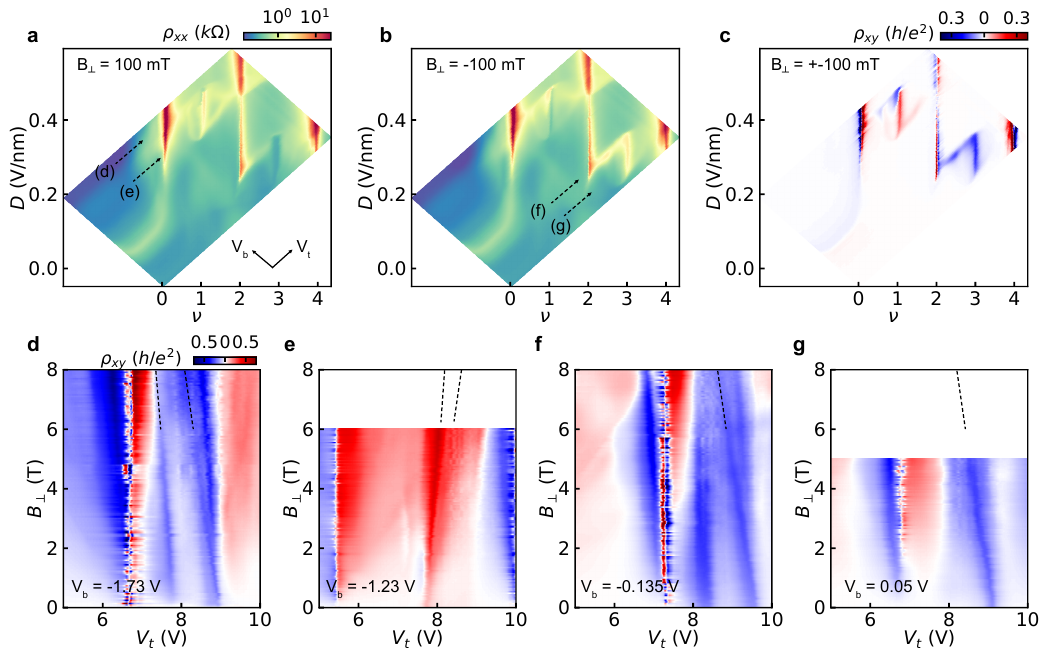} 
\caption{\textbf{Landau fans and gate maps at $\boldsymbol{T=1.7}$~K from the device at $\mathbf{\boldsymbol{\theta}=1.33^\circ}$ (D2)}. \textbf{a}, Zoomed-in gate map of $\rho_{xx}$ versus $D$ and $\nu$ taken at $B_
\perp = 100$~mT. \textbf{b}, Same as \textbf{(a)} at $B_
\perp = -100$~mT. \textbf{c}, Antisymmetrized $\rho_{xy}$ corresponding to \textbf{a} and \textbf{b}. \textbf{d}, Landau fan (antisymmetrized $\rho_{xy}$) taken along the black dashed line denoted in \textbf{a}, at $V_b = -1.73$~V, cutting through $\nu = 1/2$ and $\nu = 1$. Dashed lines correspond to $C = -1$ and $C = -2$. \textbf{e}, $V_b = -1.23$~V, cutting through $\nu = 1$. Dashed lines correspond to $C = 1$ and $C = 2$. \textbf{f}, $V_b = -0.135$~V, cutting through $\nu = 5/2$ and $\nu = 3$. Dashed line corresponds to $C = -2$. \textbf{g}, $V_b = 0.05$~V, cutting through $\nu = 3$. Dashed line corresponds to $C = -2$. All measurements are taken at $T\approx1.7$~K.
}
\label{fig:S_D2VTI}
\end{figure*}

\textbf{Anomalous hysteresis in a $\mathbf{\boldsymbol{\theta}=1.33^\circ}$ device.} In one of the $\theta=1.33^\circ$ devices, we identify features of the $\nu$--$D$ gate map that appear to track a single gate (Fig.~\ref{fig:S_D2VTI}) and exhibit electron-ratchet behavior upon reversing the gate sweeping direction. This anomalous hysteresis effect is known to occur in some graphene/hBN heterostructures~\cite{Zheng2020_MoireFerroelectricity,watersAnomalousHysteresisGraphite2025} and has been linked to the rotational alignment between the hexagonal boron nitride layers within the heterostructure~\cite{Maffione2026_AnomalousGating}, although the precise conditions leading to the behavior remain unknown. We operated this device in a way such that the anomalous hysteresis does not impact our conclusions. 

\textbf{Measurements at finite $\Bpar$.}
For measurements at finite $\Bpar$, the out-of-plane component from sample misalignment was eliminated by taking maps of $dV/dI$ versus $\Bperp$ and $I_{dc}$ (Fig.~\ref{fig:4}g,h). From these maps, true $\Bperp = 0$ was identified as the line about which the $dV/dI$ is symmetric. After this identification, we adjusted the applied values of the vector magnet to compensate for any unintentional out-of-plane field, leaving only an in-plane field applied to the sample.

\textbf{Stacking order.} The trilayer regions were not imaged during fabrication to distinguish Bernal from rhombohedral stacking. Rhombohedral domains that remain connected to Bernal regions can convert to Bernal stacking through strain-driven domain-wall motion during fabrication, and rhombohedral regions are routinely isolated by AFM- or laser-cutting before assembly~\cite{Chen2019_TLGMott}. Any rhombohedral portion of a trilayer region that also contained Bernal stacking would therefore be expected to convert during assembly. A rhombohedral trilayer would be most likely to survive assembly if the entire region isolated by local anodic oxidation happened to lie within a single rhombohedral domain, which is unlikely given that rhombohedral stacking makes up only about 15\% of the area of exfoliated trilayer flakes~\cite{Lui2011_ImagingStackingOrder}. We therefore take the trilayer in each device to be Bernal-stacked, although this was not verified directly.

\clearpage

\begin{figure*}[h]
\includegraphics[width=0.95\textwidth]{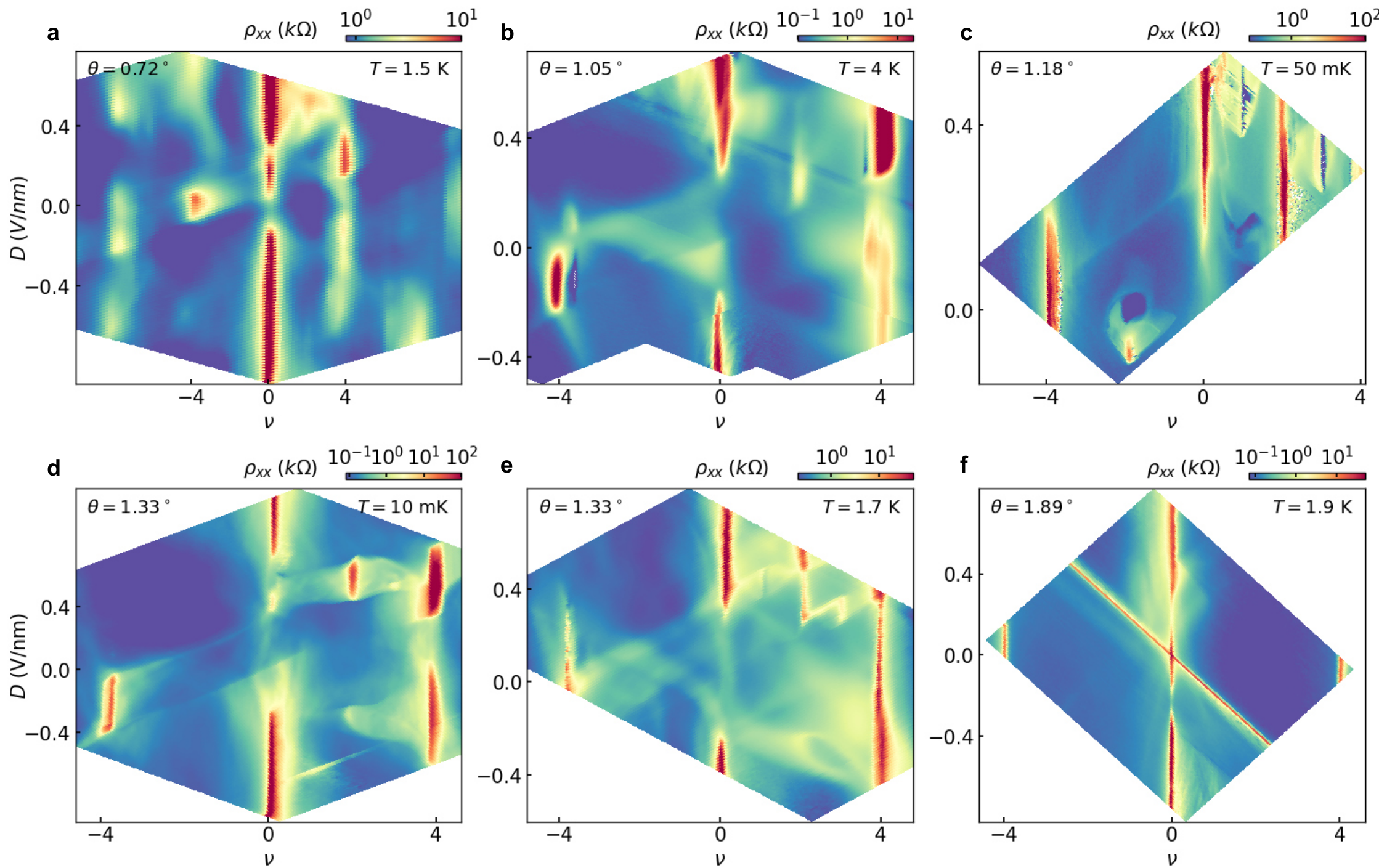} 
\caption{\textbf{Transport characterization of all devices.} $\rho_{xx}$ maps as a function of $\nu$ and D for six twisted bilayer-trilayer graphene devices with different twist angles: \textbf{a}, $\theta=0.72^\circ$ \textbf{b}, $\theta=1.05^\circ$ \textbf{c}, $\theta=1.18^\circ$ \textbf{d-e}, $\theta=1.33^\circ$ \textbf{f}, $\theta=1.89^\circ$. The calculation of twist angles is detailed in the methods section and all maps are taken at $B=0$~T.}
\label{fig:S_gatemaps}
\end{figure*}

\begin{figure*}[h]
\includegraphics[width=0.95\textwidth]{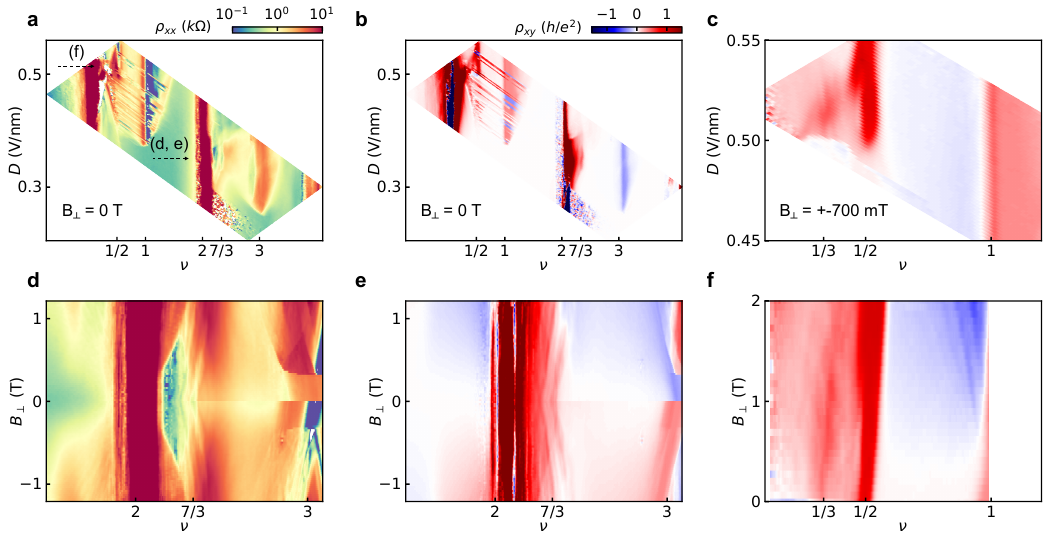} 
\caption{\textbf{Additional gate maps and Landau fans from the $\mathbf{\boldsymbol{\theta}=1.18^\circ}$ device}. \textbf{a}, Zoomed-in gate map of $\rho_{xx}$ versus $D$ and $\nu$ taken at $B = 0$~T. \textbf{b}, Same as \textbf{a} for $\rho_{xy}$. \textbf{c}, Map of the high-$D$ region of the device, taken at $B_\perp=\pm700$~mT. \textbf{d}, Landau fan taken along the black dashed line denoted in \textbf{a}. \textbf{e}, $\rho_{xy}$ corresponding to \textbf{d}. \textbf{f}, Antisymmetrized Landau fan ($\rho_{xy}$) taken at high-$D$, along the black dashed line denoted in \textbf{a}.
}
\label{fig:S_fractional}
\end{figure*}

\begin{figure*}[h]
\includegraphics[width=0.7\textwidth]{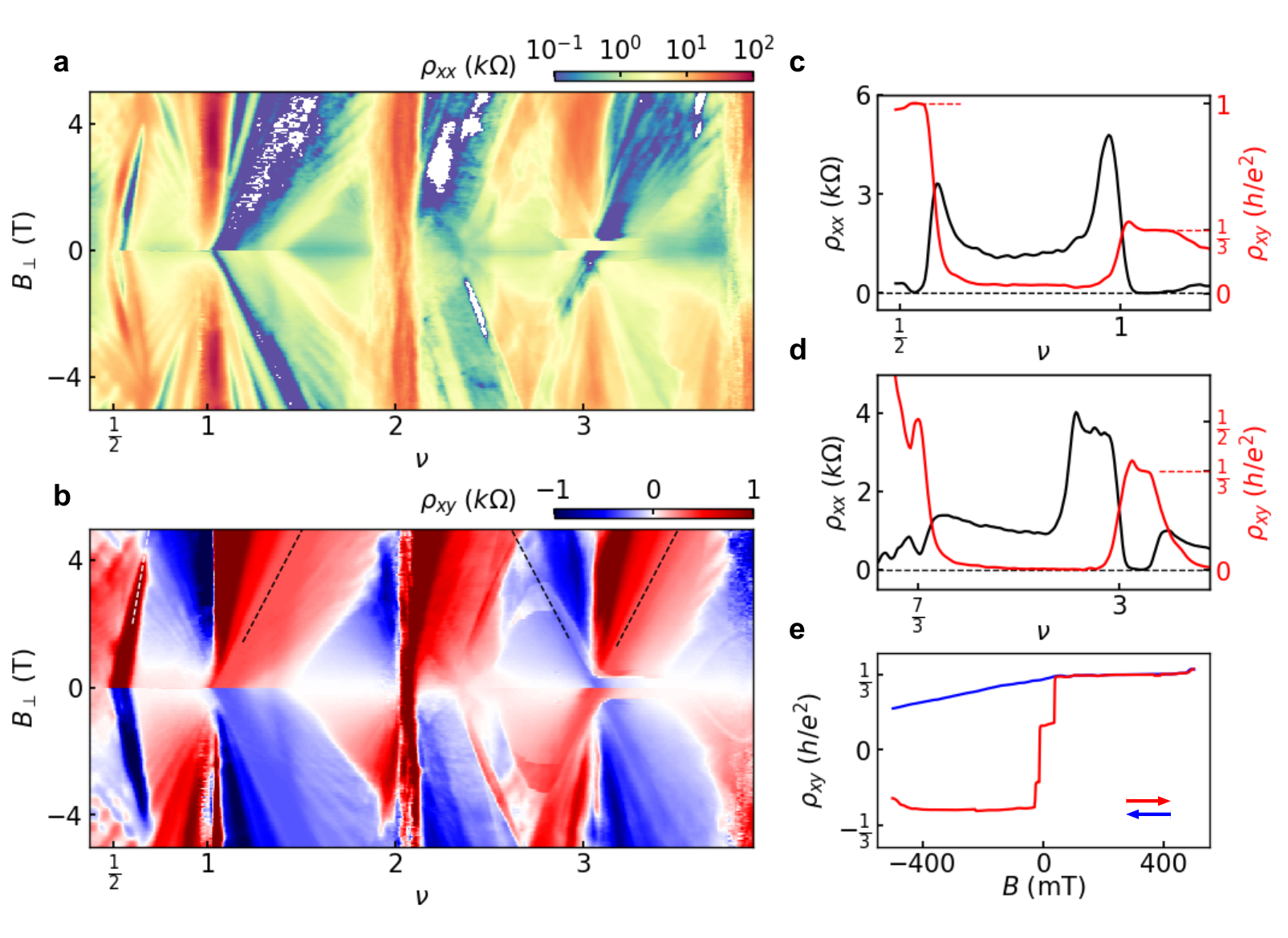} 
\caption{\textbf{High-$\boldsymbol{D}$ Landau fans and linecuts from the $\mathbf{\boldsymbol{\theta}=1.18^\circ}$ device}. \textbf{a}, Landau fans of $\rho_{xx}$ corresponding to the symmetrized fan in Fig.~\ref{fig:1}c. \textbf{b}, same for $\rho_{xy}$ corresponding to the antisymmetrized fan in Fig.~\ref{fig:1}c. White and black dashed lines indicate the Středa trajectories of C = +1, -3 and +3 states. \textbf{c}, Line cut of $\rho_{xx}$ and $\rho_{xy}$ taken along $V_{b}$ through $\nu = \frac{1}{2}$ and 1 at $V_t=9.3$~V and $B=0$~T. \textbf{d}, Linecut of $\rho_{xx}$ and $\rho_{xy}$ taken along $V_{b}$ through $\nu = 3$ at $V_{t}=8.26$~V and $B = 0$~T. \textbf{e}, $\rho_{xy}$ versus $B_\perp$ taken at $\nu = 3$ and $D=0.31$~V/nm. All measurements taken at $T=100$~mK.
}
\label{fig:S_addfans}
\end{figure*}

\begin{figure*}[h]
\includegraphics[width=0.95\textwidth]{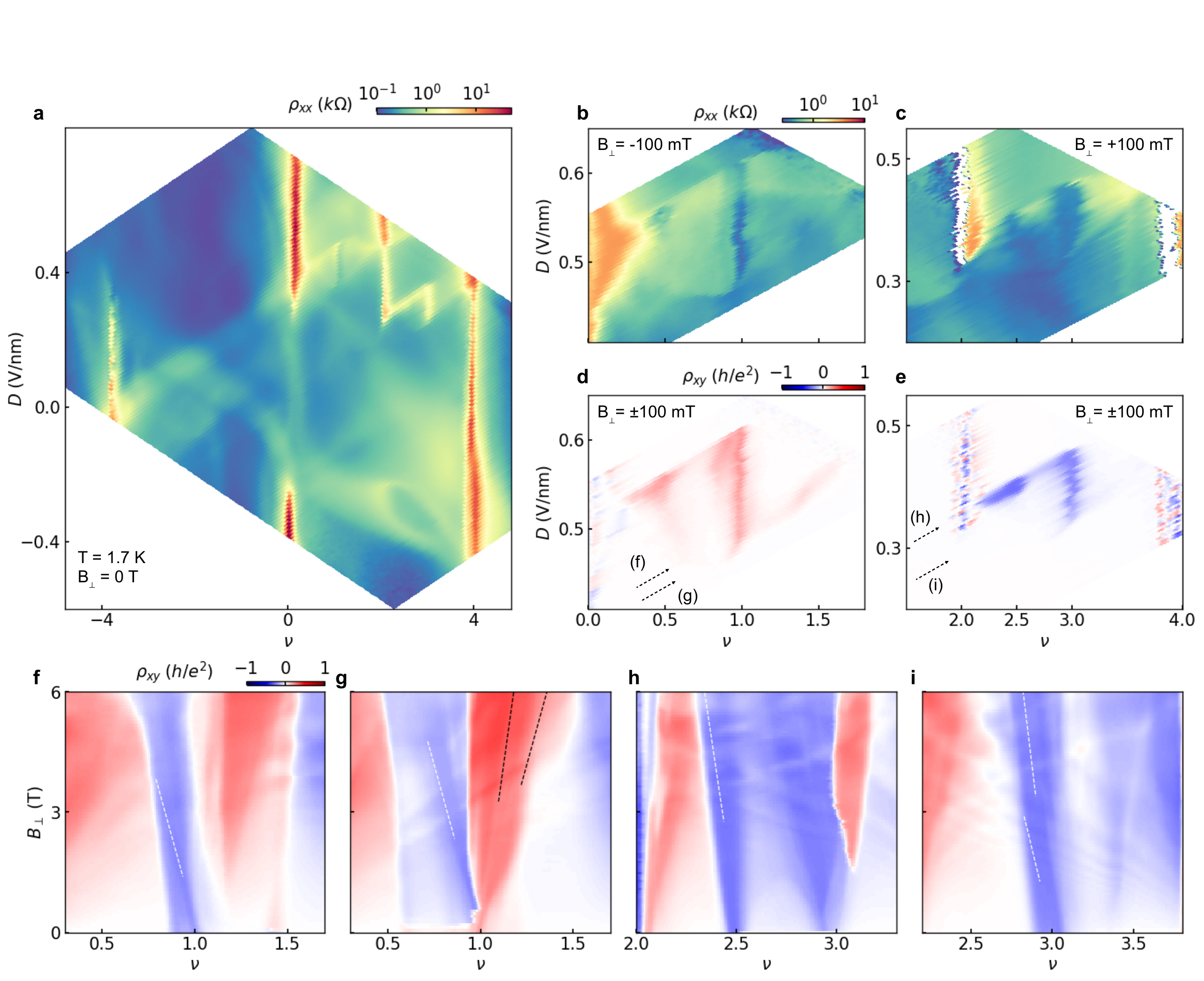} 
\caption{\textbf{Gate maps and Landau fans from the $\mathbf{\boldsymbol{\theta}=1.33^\circ}$ device (D2)}. \textbf{a}, Map of $\rho_{xx}$ versus $D$ and $\nu$ taken over the full gate range at $T = 1.7$~K. \textbf{b}, Zoomed-in map of $\rho_{xx}$ near $\nu = 1$ taken at $T = 10$~mK and $B_\perp = -100$~mT. \textbf{c}, Similar map to \textbf{b}, taken around $\nu = 3$ and $B_\perp=100$~mT. \textbf{d, e}, $\rho_{xy}$ (antisymmetrized), corresponding to \textbf{b, c}. \textbf{f-i}, Landau fans of $\rho_{xy}$ (antisymmetrized) taken at \textbf{f}, $V_t = 7$~V; \textbf{g}, $V_t = 7.2$~V; \textbf{h}, $V_t = 4.8$~V; \textbf{i}, $V_t = 7.2$~V. White and black dashed lines indicate the St\v{r}eda trajectories of C = -1,+1, -2, and +2 states. Note that in fans \textbf{g} and \textbf{i} the magnitude of the Chern number at $B=0$~T was ambiguous between $|C|=1$ and $|C|=2$.}
\label{fig:S_D2fans}
\end{figure*}

\begin{figure*}[h]
\includegraphics[width=0.95\textwidth]{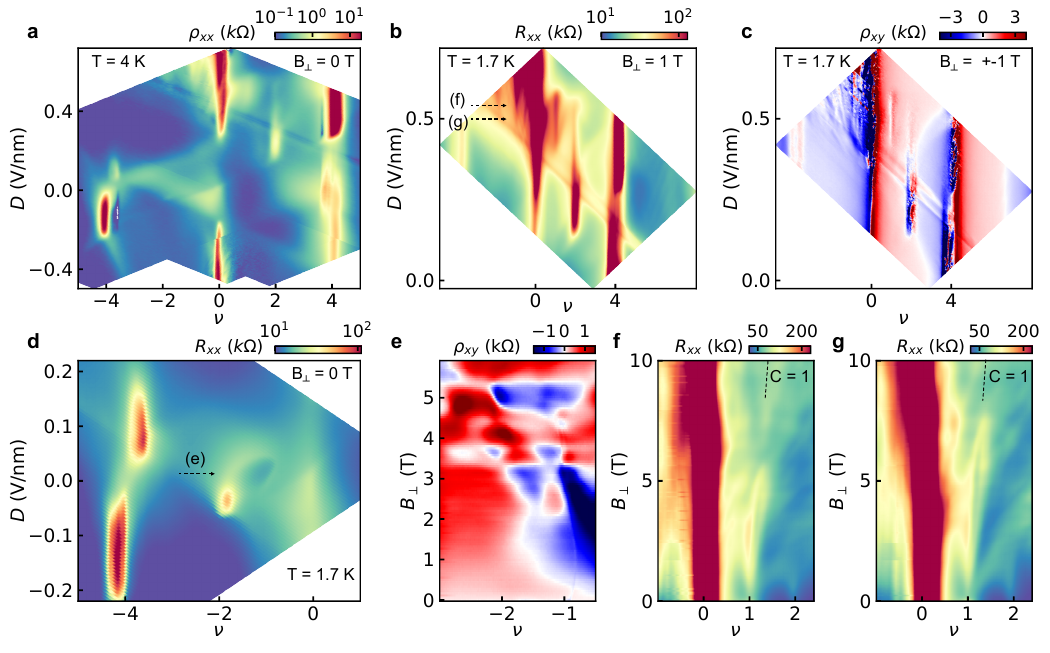} 
\caption{\textbf{Measurements from the $\mathbf{\boldsymbol{\theta}=1.05^\circ}$ device (D3)}. \textbf{a}, Map of $\rho_{xx}$ versus $D$ and $\nu$ over the full gate range at $T = 4$~K. \textbf{b}, Zoom in map of the high-$D$ region taken at $T = 1.7$~K and $B = 1$~T. \textbf{c}, $\rho_{xy}$ (antisymmetrized) corresponding to \textbf{b}. \textbf{d}, Zoom in map of the hole doping side of the phase diagram showing $R_{xx}$ versus $D$ and $\nu$. \textbf{e}, Landau fan showing antisymmetrized $\rho_{xy}$, taken along the black dashed line in \textbf{d} at $D \approx 0.015$~V/nm. \textbf{f, g} Landau fan showing $R_{xx}$, taken along the dashed lines denoted in \textbf{b} ($D \approx 0.53$~V/nm, $D \approx 0.49$~V/nm). Measurements labeled with $R_{xx}$ were taken in a 3-point contact configuration. Measurements labeled with $\rho_{xx}$ were taken in a 4-point contact configuration. 
}
\label{fig:S_1degree}
\end{figure*}

\begin{figure*}[h]
\includegraphics[width=0.95\textwidth]{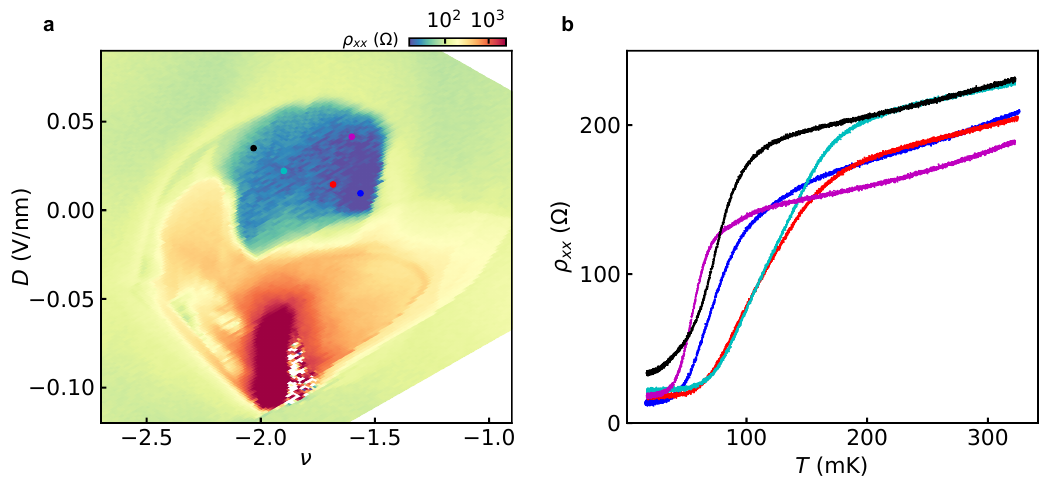} 
\caption{\textbf{$\boldsymbol{\rho_{xx}}$ versus $T$ in the low resistance pocket}. \textbf{a}, Map of $\rho_{xx}$ versus $\nu$ and $D$ in the region surrounding the correlated halo at $\nu = -2$ (Device D1). \textbf{b}, $\rho_{xx}$ versus $T$ curves taken at different points within the low resistance pocket. Curve colors match the dots in \textbf{a}.
}
\label{fig:S_SC1DRT}
\end{figure*}

\begin{figure*}[h]
\includegraphics[width=0.95\textwidth]{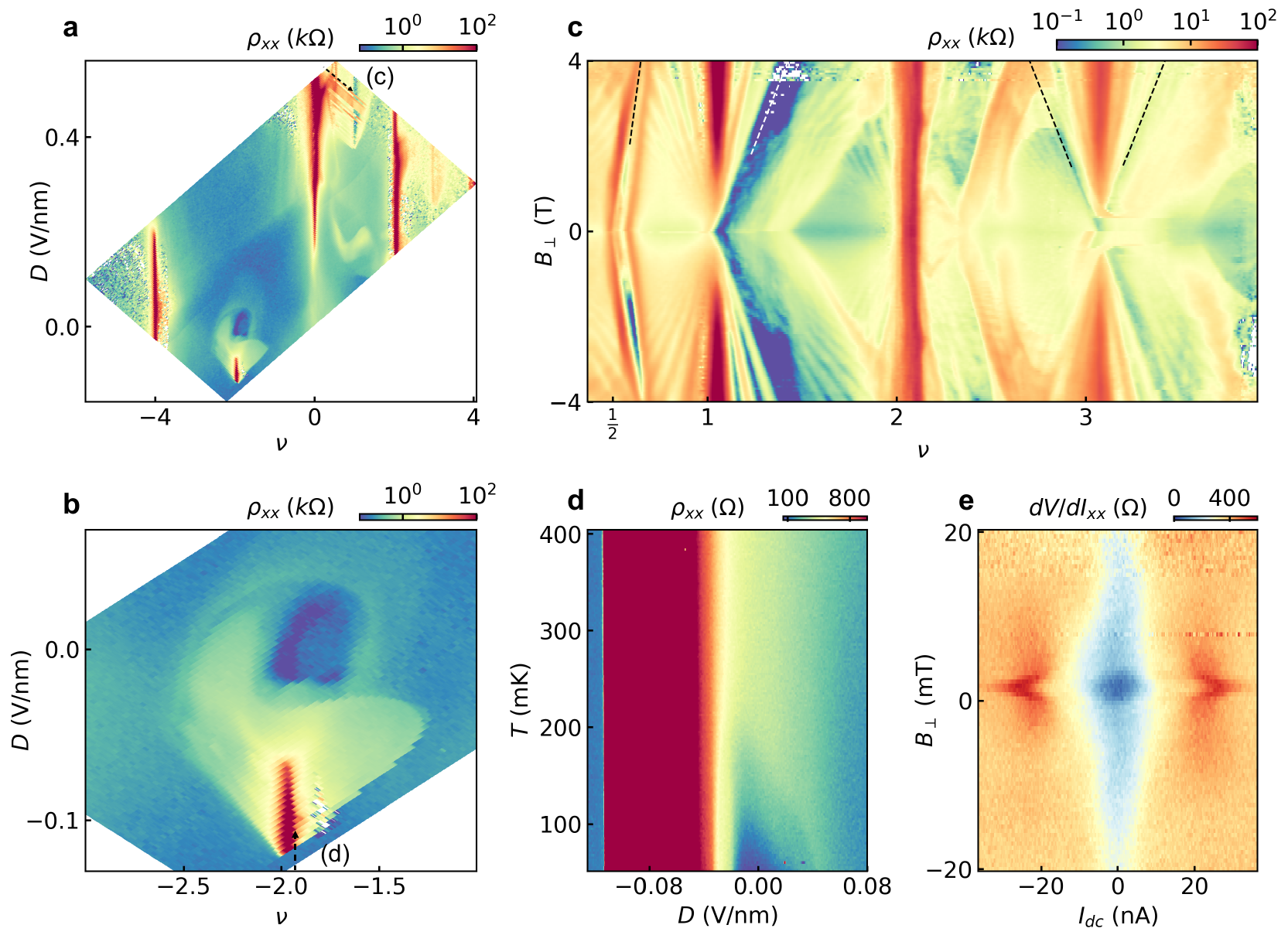} 
\caption{\textbf{Measurements from a second contact pair in the $\mathbf{\boldsymbol{\theta}=1.18^\circ}$ device}. \textbf{a}, Map of $\rho_{xx}$ versus $D$ and $\nu$. \textbf{b}, Zoomed-in map of $\rho_{xx}$ in the halo region near $\nu = -2$. \textbf{c}, Landau fan taken at $V_t = 9.05$~V (same as Fig.~\ref{fig:S_addfans}a) denoted by the black dashed line in \textbf{a}. White and black dashed lines indicate the St\v{r}eda trajectories of $C$ = $+1$, $+3$, and $-3$ states. \textbf{d}, $\rho_{xx}$ versus $T$ and $D$ taken at $\nu = -1.95$ denoted by the black dashed line in \textbf{b}. \textbf{e}, $dV/dI$ versus $I_{dc}$ and $B_\perp$ taken at $\nu=-1.95$ and $D=0.01$~V/nm. All measurements, except \textbf{d}, performed at $T=50$~mK.
}
\label{fig:S_D1contactpair2}
\end{figure*}

\begin{figure*}[h]
\includegraphics[width=0.95\textwidth]{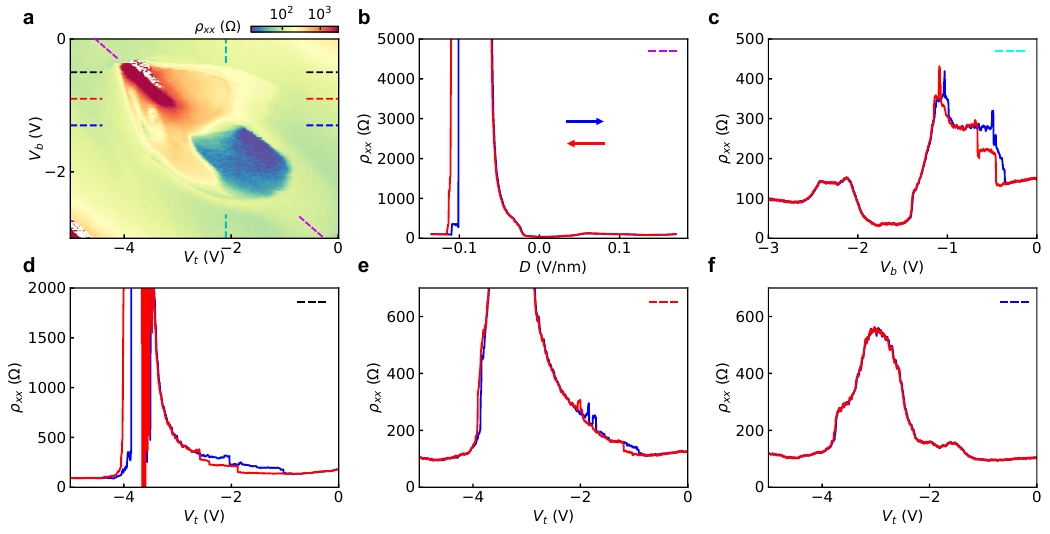} 
\caption{\textbf{Hysteresis across the halo boundary}. \textbf{a}, Gate map of $\rho_{xx}$ versus $V_b$ and $V_t$ (Device D1). \textbf{b}, Forward (blue) and backward (red) sweeps taken at constant $\nu = -1.95$ along the dashed line denoted in \textbf{a}. \textbf{c}, Same as \textbf{b}, at constant $V_t = -2.1$~V. \textbf{d}, $V_b = -0.5$~V. \textbf{e}, $V_b = -0.9$~V. \textbf{f}, $V_b = -1.3$~V.
}
\label{fig:S_Hysteresis}
\end{figure*}

\begin{figure*}[h]
\includegraphics[width=0.95\textwidth]{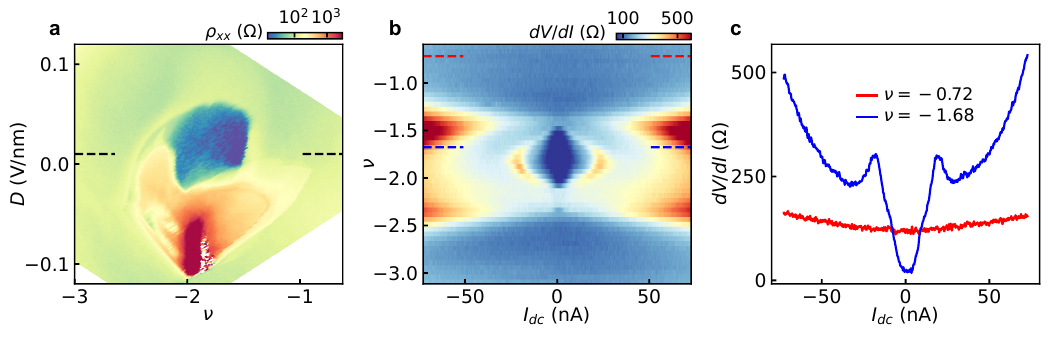} 
\caption{\textbf{$\boldsymbol{dV/dI}$ versus $\nu$ across the low resistance pocket}. \textbf{a}, Map of $\rho_{xx}$ versus $\nu$ and $D$ in the region surrounding the correlated halo at $\nu = -2$ (Device D1). \textbf{b}, $dV/dI$ versus $\nu$ and $I_{dc}$ taken at constant $D = 0.01$~V/nm (black dashed line in \textbf{a}) across the low resistance pocket. \textbf{c}, Linecuts taken along the blue and red dashed lines in \textbf{b}.
}
\label{fig:S_dVdIconstD}
\end{figure*}

\begin{figure*}[h]
\includegraphics[width=0.6\textwidth]{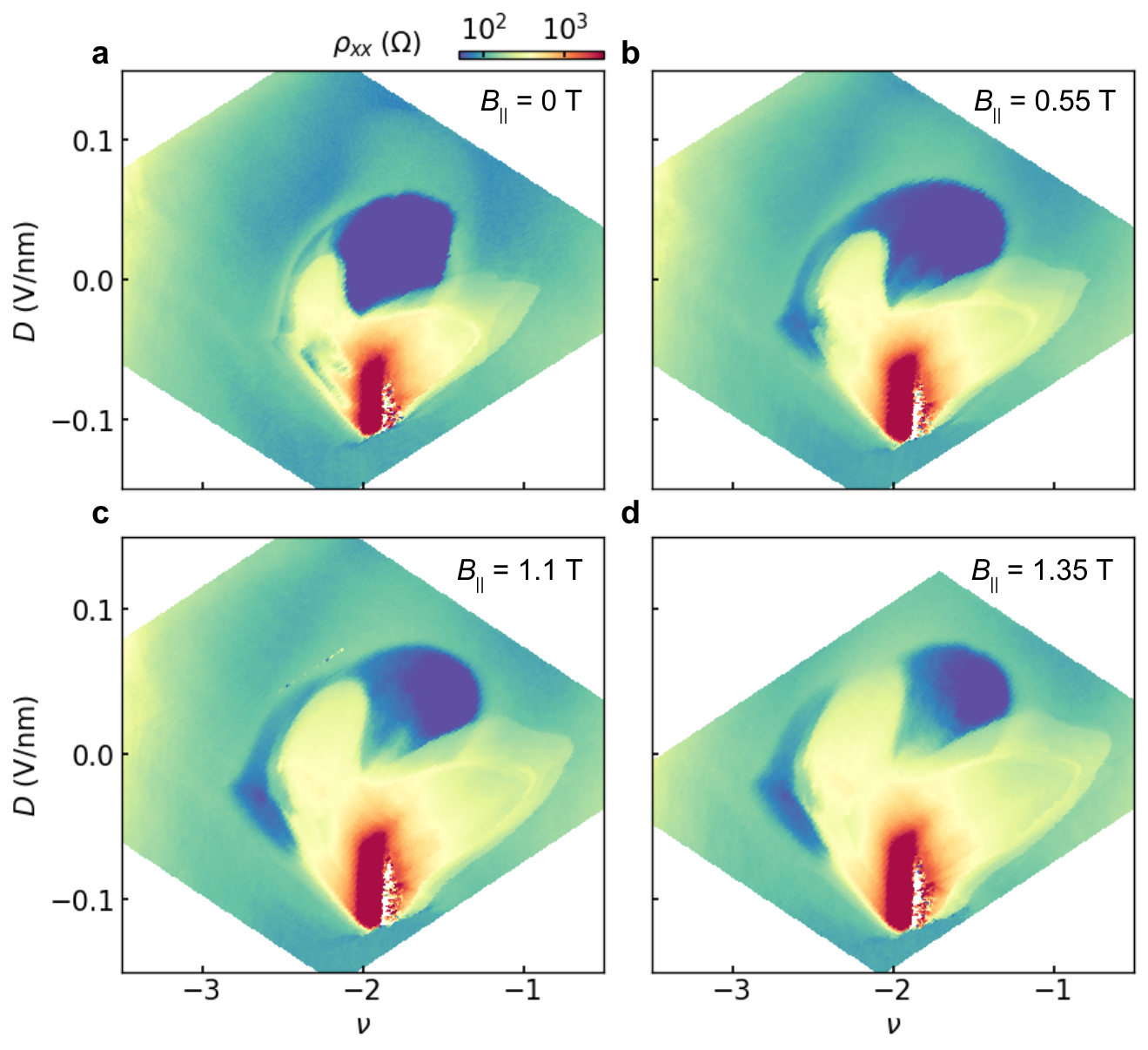} 
\caption{\textbf{Parallel field dependence of the halo region.} \textbf{a}, Map of $\rho_{xx}$ versus $\nu$ and $D$ at $B_\parallel=0$~T in the region surrounding the correlated halo at $\nu = -2$. \textbf{b}, Same as \textbf{a} for $B_\parallel=0.55$~T. \textbf{c}, $B_\parallel=1.1$~T. \textbf{d}, $B_\parallel=1.35$~T. All maps correspond to the schematic in Fig. \ref{fig:4}b and were taken at $B_\perp=0$~T.}
\label{fig:S_BparHaloEvolution}
\end{figure*}

\begin{figure*}[h]
\includegraphics[width=0.95\textwidth]{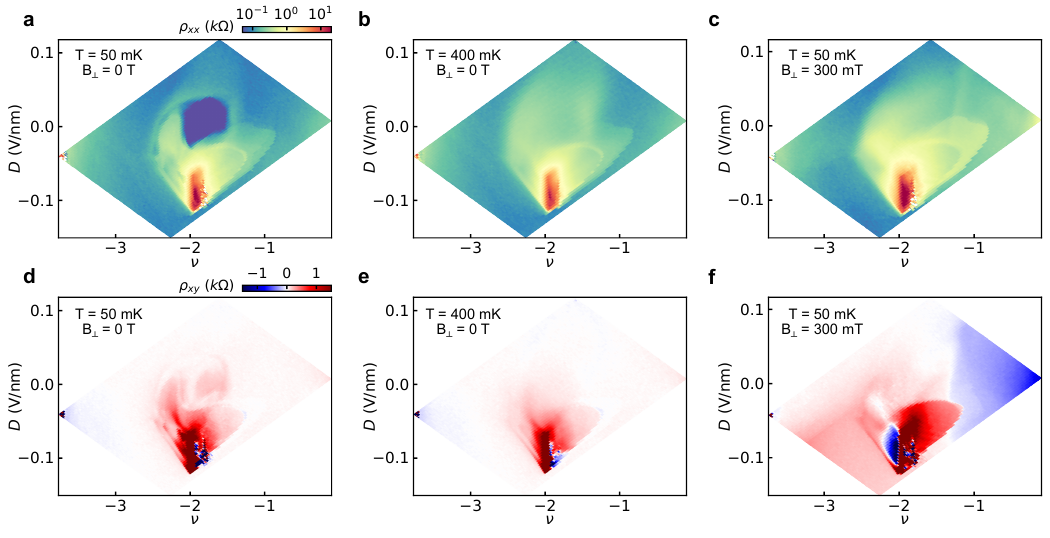} 
\caption{\textbf{Comparison of the halo region at different temperatures and magnetic fields}. \textbf{a}, Map of $\rho_{xx}$ versus $D$ and $\nu$ taken at $T = 50$~mK and $B = 0$. \textbf{b}, Same as \textbf{a}, taken at $T = 400$~mK. \textbf{c} $B = 300$~mT. \textbf{d-f}, $\rho_{xy}$ corresponding to \textbf{a-c}. Measurements are taken in the Bluefors LD dilution refrigerator with a one-axis superconducting magnet.
}
\label{fig:S_TandBdepHalo}
\end{figure*}

\begin{figure*}[h]
\includegraphics[width=0.95\textwidth]{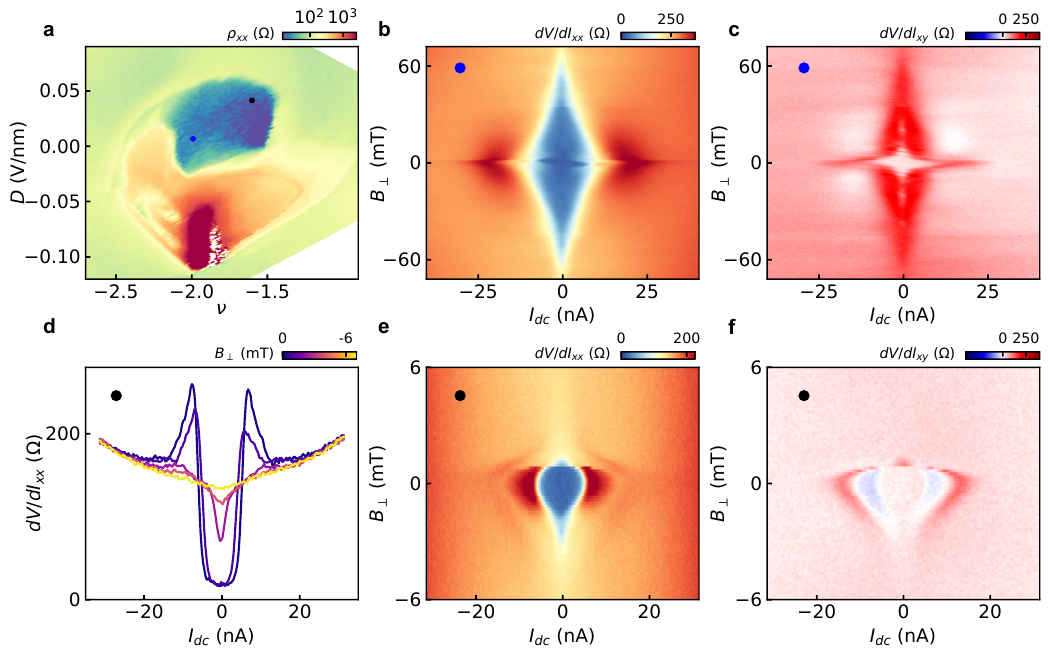} 
\caption{\textbf{Differential resistance measurements in the incipient superconducting state}. \textbf{a}, Map of $\rho_{xx}$ versus $\nu$ and $D$ in the region surrounding the correlated halo at $\nu = -2$ (Device D1). \textbf{b}, $dV/dI_{xx}$ versus $B_\perp$ and $I_{dc}$, taken at the blue dot in \textbf{a}. \textbf{c}, $dV/dI_{xy}$ corresponding to \textbf{b}. \textbf{d}, Linecuts of $dV/dI_{xx}$ at different $B_\perp$, taken from the map in \textbf{e}. \textbf{e}, Same as \textbf{b}, taken at the black dot in \textbf{a}. \textbf{f}, $dV/dI_{xy}$ corresponding to \textbf{e}.
}
\label{fig:S_SCdVdI}
\end{figure*}

\begin{figure*}[h]
\includegraphics[width=0.95\textwidth]{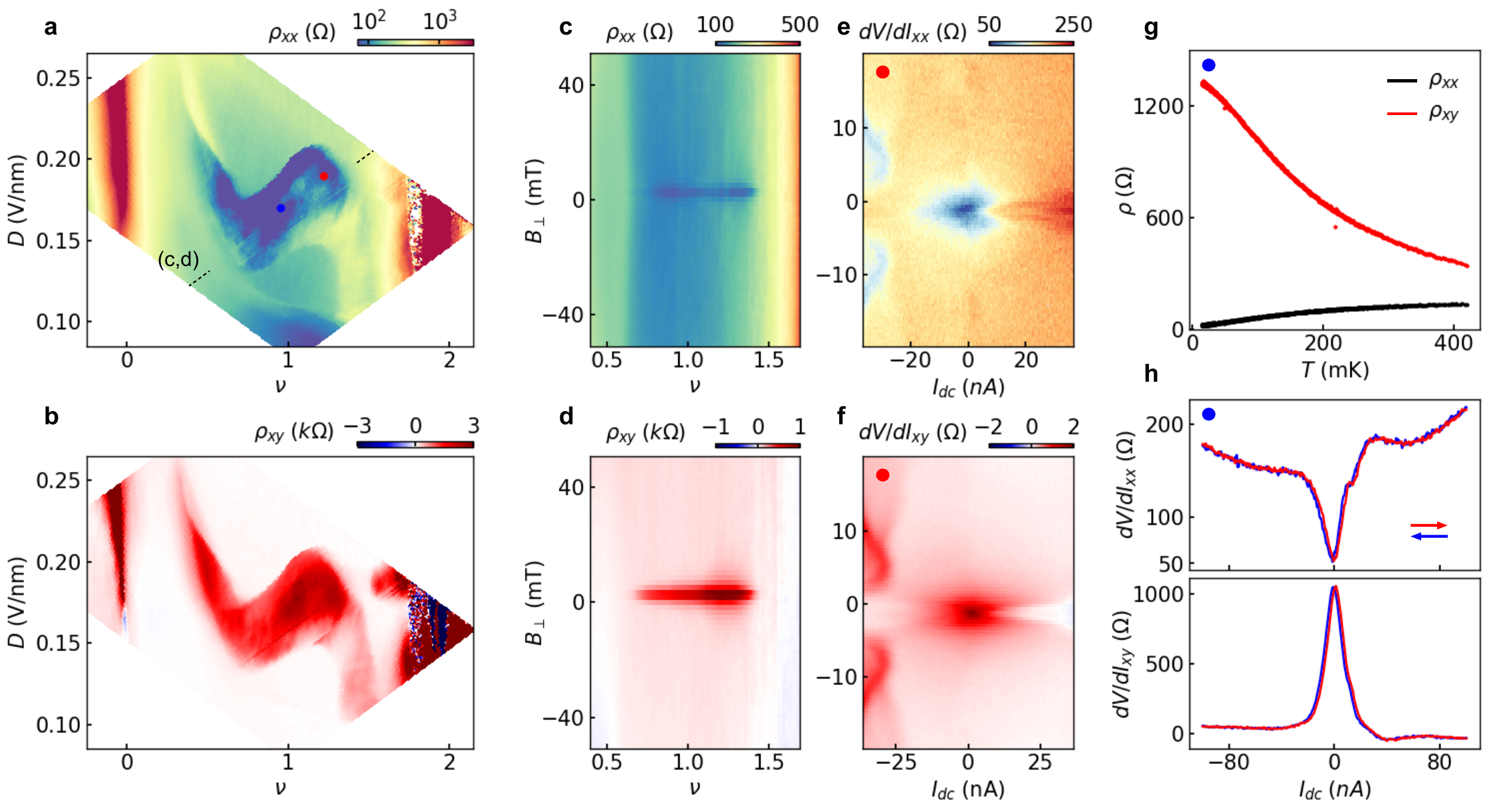} 
\caption{\textbf{Analysis of the low-resistance region for electron doping}. \textbf{a}, Map of $\rho_{xx}$ versus $D$ and $\nu$, zoomed-in around the low-resistance region at electron dopings and $B=0$~T. \textbf{b}, Same as \textbf{a} for $\rho_{xy}$. \textbf{c}, $\rho_{xx}$ versus $B_\perp$ and $\nu$ along the dashed line shown in \textbf{a}, taken at constant $V_b = -1.2$~V. \textbf{d}, Same as \textbf{c} for $\rho_{xy}$. \textbf{e}, $dV/dI_{xx}$ versus $I_{dc}$ and $B_\perp$, taken at the red dot in \textbf{a}, $\nu =1.22$, $D =0.19 $~V/nm. \textbf{f}, Same as \textbf{e} for $dV/dI_{xy}$. \textbf{g}, 1D linetrace of $\rho_{xx}$ and $\rho_{xy}$ taken at the blue dot in \textbf{a}, $\nu =0.95$, $D =0.17 $~V/nm. \textbf{h}, Line cuts of $dV/dI_{xx}$ and $dV/dI_{xy}$ taken at the blue dot in \textbf{a}.
}

\label{fig:S_anisotropic}
\end{figure*}

\begin{figure*}[h]
\includegraphics[width=0.50\textwidth]{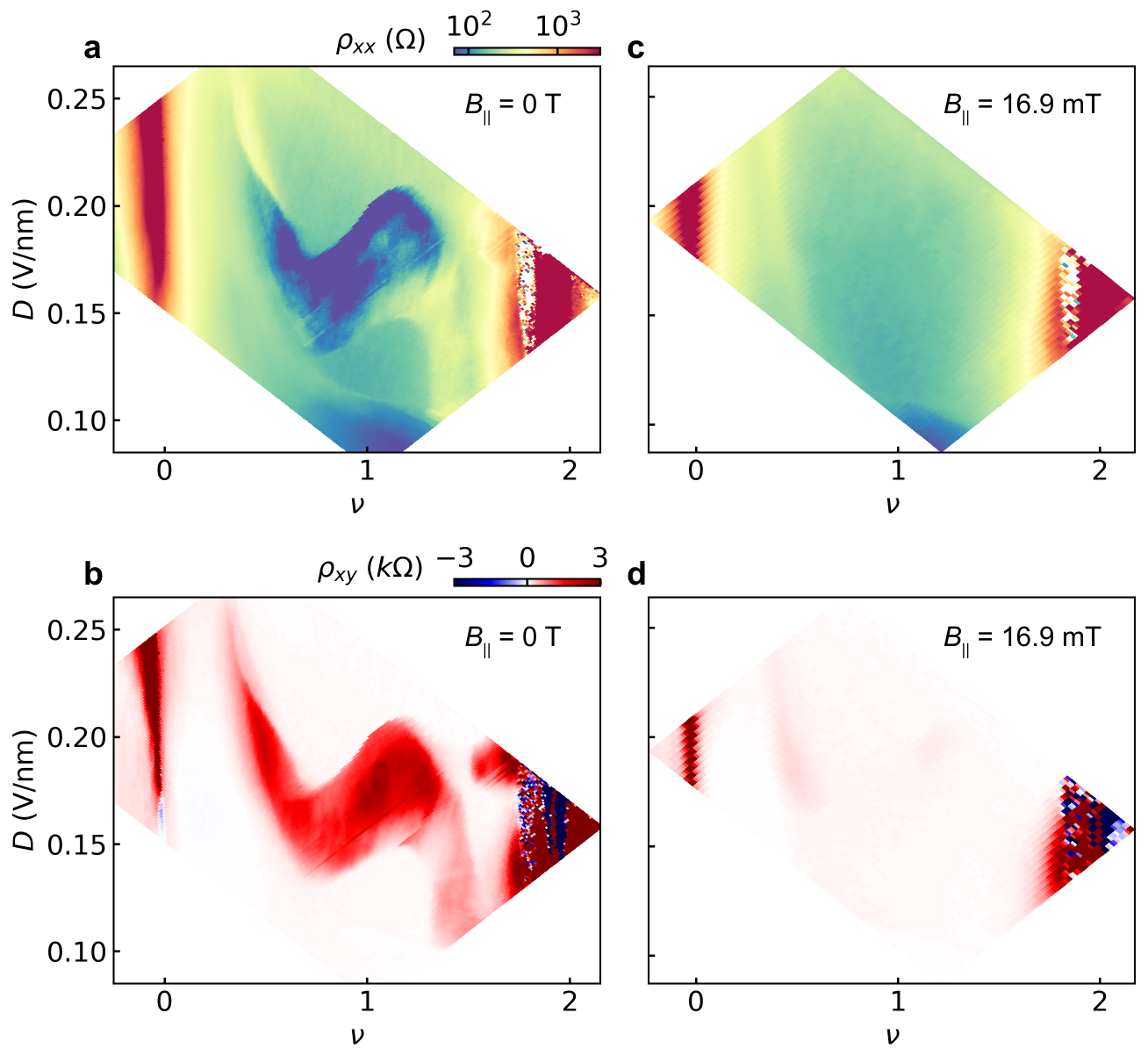} 
\caption{\textbf{Parallel field dependence of the low-resistance region for electron doping} \textbf{a,} Map of $\rho_{xx}$ versus $D$ and $\nu$, zoomed-in around the low-resistance region at electron dopings and $B=0$~T (same as Fig.~\ref{fig:S_anisotropic}a). \textbf{b,} Same as \textbf{a} for $\rho_{xy}$ (same as Fig.~\ref{fig:S_anisotropic}b). \textbf{c,} Same as \textbf{a} taken at $B_\parallel=16.9$~mT. \textbf{d,} Same as \textbf{c} for $\rho_{xy}$. All measurements taken at $T=8$~mK and $B_\perp=0$~T.
}
\label{fig:S_anisotropic_BparSensisitivity}
\end{figure*}

\begin{table}[H]
\centering
\renewcommand{\arraystretch}{2.5}
\setlength{\tabcolsep}{3pt}
\begin{tabular}{|c|c|c|c|c|c|c|c|c|c|c|c|}
\hline
$\boldsymbol{\theta}$ \textbf{(degrees)} & $\boldsymbol{\nu = -2}$ & $\boldsymbol{\nu = \frac{1}{4}}$ & $\boldsymbol{\nu = \frac{1}{3}}$ & $\boldsymbol{\nu = \frac{1}{2}}$ & $\boldsymbol{\nu = \frac{2}{3}}$ & $\boldsymbol{\nu = 1}$ & $\boldsymbol{\nu = \frac{3}{2}}$ & $\boldsymbol{\nu = 2}$ & $\boldsymbol{\nu = \frac{7}{3}}$ & $\boldsymbol{\nu = \frac{5}{2}}$ & $\boldsymbol{\nu = 3}$ \\
\hline\hline
$1.50^\circ$ & $C=0$ & $|C|=1$ & $|C|=1^\dagger$ & $|C|=1^\dagger$ & $|C|=1^\dagger$ & $C=0$ & $|C|=1^\dagger$ & $C=0$ & & & $C=0$ \\
\hline
$1.33^\circ$ & & & & \cellcolor{gray!10}$|C|=1^\dagger$ & & \cellcolor{gray!10}$|C|=2^{\diamond\:?}$ & & $C=0$ & & \cellcolor{gray!10}$|C|=1$ & \cellcolor{gray!10}$|C|=2^?$ \\
\hline
$1.18^\circ$ & $C=0$ & & \cellcolor{gray!10}$|C|=1^?$ & $|C|=1$ & & $|C|=3$ & & $C=0$ & \cellcolor{gray!10}$|C|=2^?$ & & $|C|=3^\diamond$ \\
\hline
$1.05^\circ$ & CI & & & & & \cellcolor{gray!10}$|C|=1^{\dagger}$ & & $C=0$ & & & \\
\hline
$0.72^\circ$ & & & & & & $C=0$ & & $C=0$ & & & \\
\hline
\end{tabular}
\vspace{4pt}
\begin{minipage}{\linewidth}
\footnotesize
\raggedright
$^\dagger$ Indicates states that form at modest finite field \\
$^\diamond$ Indicates that both the positive and negative Chern states are present in the Landau fan diagrams\\
$^?$ Indicates states where the Chern number is not clearly identified. See caption for further discussion. \\
Grey shading indicates an underdeveloped state with $|C|>0$ whose Chern number is inferred from the Landau fan diagrams. \\
CI denotes a correlated insulator whose behavior with $B$ was not studied \\
Empty cell denotes metallic state
\end{minipage}
\caption{\textbf{Summary of the observed Chern numbers across all devices.} Chern numbers of correlated insulating states at different $\nu$ for devices of varying twist angle $\theta$. Note that for the $\nu=\frac{1}{3}$ and $\nu=\frac{7}{3}$ states in the $\theta=1.18^\circ$ device, we are not able to distinguish between an integer or fractional state. For the $\nu=1$ and $\nu=3$ states in the $\theta=1.33^\circ$ device, it is also difficult to distinguish between $|C|=1$ or 2 at $B=0$~T (see Fig.~\ref{fig:S_D2fans} and Fig.~\ref{fig:S_D2VTI}), although we note that at $\nu=1$ there is a clear $|C|=2$ feature that travels to zero field. Data from the $\theta = 1.50^\circ$ device is from Ref.~\cite{Su2025}.}
\label{tab:chern_numbers}
\end{table}

\end{document}